\documentclass[manuscript]{aastex61}

\newcommand\aastex{AAS\TeX}

\shorttitle{\aastex\ Dynamical TS}
\shortauthors{Shen et al.}

\begin{document}

%
%
\title{The Dynamical Behavior of Reconnection-driven Termination Shocks in Solar Flares: Magnetohydrodynamic Simulations}

\author[0000-0002-9258-4490]{Chengcai Shen}
\affiliation{Harvard-Smithsonian Center for Astrophysics, 60, Garden Street, Cambridge, MA, 02138, USA}
\author[0000-0003-1034-5857]{Xiangliang Kong}
\affiliation{Shandong Provincial Key Laboratory of Optical Astronomy and Solar-Terrestrial Environment,
and Institute of Space Sciences, Shandong University, Weihai, Shandong 264209, China}
\author[0000-0003-4315-3755]{Fan Guo}
\affiliation{Los Alamos National Laboratory, P.O. Box 1663, Los Alamos, NM 87545, USA}
\author[0000-0002-7868-1622]{John C. Raymond}
\affiliation{Harvard-Smithsonian Center for Astrophysics, 60, Garden Street, Cambridge, MA, 02138, USA}
\author[0000-0002-0660-3350]{Bin Chen}
\affiliation{Center for Solar-Terrestrial Research, New Jersey Institute of Technology,
323 Dr. Martin Luther King Blvd, Newark, NJ 07102, USA}

\begin{abstract}
In eruptive solar flares, termination shocks (TSs), formed when high-speed reconnection outflows collide with closed dense flaring loops, are believed to be one of the possible candidates for plasma heating and particle acceleration. 
In this work, we perform resistive magnetohydrodynamic simulations in a classic Kopp-Pneuman flare configuration to study the formation and evolution of TSs, and analyze in detail the dynamic features of TSs and variations of the shock strength in space and time. 
This research focuses on the fast reconnection phase when plasmoids form and produce small-scale structures inside the flare current sheet.
It is found that the TS emerges once the downward outflow colliding with closed magnetic loops becomes super-magnetosonic, and immediately becomes highly dynamical.
The morphology of a TS can be flat, oblique, or curved depending on the detailed interactions between the outflows/plasmoids and the highly dynamic plasma in the looptop region. 
%
%
The TS becomes weaker when a plasmoid is crossing through, or may even be destroyed by well developed plasmoids and then re-constructed above the plasmoids.
We also perform detailed statistical analysis on important physical quantities along and across the shock front.
The density and temperature ratios range from 1 to 3 across the TS front, and the pressure ratio typically has larger values up to 10.
We show that weak guide fields do not strongly affect the Mach number and compression ratios, and the TS length becomes slightly larger in the case with thermal conduction.

\end{abstract}


\onecolumngrid

\section{Introduction}
It has been widely accepted that a huge amount of magnetic energy (up to 10$^{32}$ erg) is violently released via magnetic reconnection in a typical solar eruption. The released magnetic energy is quickly converted into bulk plasma kinetic energy, plasma thermal energy, and energy contained in accelerated energetic particles.
However, the exact mechanism for accelerating the charged particles remains unclear.
There are several competing mechanisms for explaining the acceleration of charged particles in solar flares, including magnetic reconnection electric currents, turbulence/waves, and fast-mode shocks \citep[see][and references therein]{Miller1997,Zharkova2011}. 

A termination shock (TS) has been predicted in early solar flare models. In the standard flare model, also known as the CSHKP model \citep{Carmichael1964,Sturrock1968,Hirayama1974,Kopp1976}, fast reconnection outflow jets are produced as a result of magnetic reconnection. They collide with the newly closed magnetic loops and 
may form a fast-mode TS, if the outflow speed exceeds the local fast-magnetosonic speed in the looptop region \citep{Forbes1986}. 
The flare TS has been suggested as a possible particle accelerator by many authors \citep{Masuda1994,Shibata1995,Forbes1996,Somov1997,Tsuneta1998,Mann2009,Guo2012,Kong2013,Li2013}, and is widely adopted in standard flare model cartoons including the arguably most famous one by \citet{Shibata1995}. It may also play an important role in the heating of plasma in or above the post-reconnection flare loops
\citep[e.g.,][]{Masuda1994,Guidoni2015} . 
%
%

The TSs also have been investigated in numerical studies
\citep[e.g.,][]{Forbes1986,Forbes1988,Workman2011,Takasao2015}. 
A stationary fast shock is identified in magnetohydrodynamic (MHD) numerical experiments with line-tied magnetic reconnection when the downward-directed reconnection jet encounters the obstacle formed by the closed loops \citep{Forbes1986a}.  
In general, the shock normal of the flare TS is nearly perpendicular to the magnetic field (i.e., the angle between the upstream magnetic field and shock normal vector $\theta_{Bn} > 45^{\circ}$). 
\citet{Forbes1986a} found the existence of this shock with a compression ratio of 2.0 and a Mach number as high as 2.3. The numerical simulations by \citet{Workman2011} have shown similar results.
%
%
\citet{Forbes1986a} also pointed out that the transition from the supermagnetosonic flow region upstream of the shock to the nearly static downstream is complicated, and a deflection sheath downstream of the fast shock is necessary for the formation of fast shocks according to the MHD jump conditions.
More recently, structures of TSs have received more attention in MHD simulations by \citet{Takasao2015} and \citet{Takasao2016}.
By including essential physics for solar flares such as magnetic reconnection, heat conduction, and chromospheric evaporation, their numerical model revealed that flare loops and the above-the-loop-top region are filled with various shocks and waves. They reported multiple shocks, including horizontal and oblique shocks above the looptop, 
and found the quasi-periodically oscillations in the above-the-loop-top regions.
This suggests that the structure of TSs is more complex than previously assumed and could significantly affect energetic electron acceleration and plasma heating, and the associated observational signatures.
Although TSs are often invoked in the standard solar eruption models, there are few solid observational constraints because they are difficult to observe. One piece of evidence for TSs is the looptop radio emission, which shows spectroscopic features similar to solar type II radio bursts (an indication of shocks in the corona) but with small frequency drift rates 
\citep{Aurass2002, Aurass2004}.
If these radio bursts are associated with plasma radiation for which the emission frequency  $f \propto n_e^{1/2}$ (where $n_e$ is the plasma density), a small frequency drift rate suggests a slow overall change of plasma density during the shock evolution, which implies a quasi-standing shock wave located in the looptop region. 
%
%
Recently, using the Karl G. Jansky Very Large Array (VLA), \citet{Chen2015} presented radio spectroscopic imaging of an eruptive solar flare event. Their imaging results showed that a TS, which appeared in the radio dynamic spectrum as a slow-drift type-II-burst-like feature,  is located in the looptop region in front of fast reconnection outflows. The corresponding RHESSI observation shows that the HXR loop-top source is nearly co-spatial with, but slightly below the shock front delineated by the radio source centroids made at different frequencies, agreeing well with the scenario that the TS is responsible for accelerating high-energy electrons and produces HXR emission in the shock downstream region. Furthermore, radio spectroscopic observations have suggested that the shock is unsteady in nature based on small but nonzero frequency drift in the dynamic spectrum (Aurass et al. 2002, 2004). This has been confirmed by the radio imaging studies in \citet{Chen2015}, which showed that the shock front revealed by radio imaging is indeed highly variable in time. 

Chen et al's observations also show the disruption and restoration of the TS possibly caused by intermittent reconnection. They reported that the TS front, as delineated by the radio source centroids, reacts dynamically to the arrival of fast plasma downflows. They showed an example in which the TS front can deviate from its initially close-to-flat geometry and turn into a concave shape during the arrival of a plasma downflow. Shortly after the interaction the shock front may be restored to its original state. Their accompanying MHD experiment results (discussed in detail in their Supplementary Materials) show that the TS morphology can be modified either upward or downward, broadened or narrowed (in its horizontal extent) in response to the arrival of the upstream magnetic structures and fast plasma flows. Sometimes the TS signature can temporarily disappear from the loop-top region.
%
%

Although the simulation in \citet{Chen2015} has shown that dynamic features of TSs can be caused by intermittent reconnection and associated small structures, the formation conditions of these dynamical TSs and their physical properties deem further investigation. In addition, thermal conduction and guide field were not included in their simulations.
In this study, we carry out a detailed investigation regarding the formation, strength, and dynamics of flare TSs, as well as the effects of introducing a guide field and thermal conduction on the shock formation and evolution.  We describe the numerical model setup in Section 2. The numerical results are presented in Section 3. 
Discussions of implications of the results are given in Section 4, and conclusions are given in Section 5.

\section{Models}
\subsection{Overview}
Our MHD model follows the classic two-ribbon flare configuration in two dimensions. It has been used to study the evolution of small-scale structures (e.g., plasmoids) inside the reconnecting current sheet, the evolution of magnetic loop and magnetic arch structures (Shen et al., 2011, 2013a), and the dynamic features of TSs above flare loops (Chen et al., 2015). The simulation starts with a current sheet in mechanical and thermal equilibrium that separates two regions of magnetic field with opposite polarity. By adding a small perturbation on the initial equilibrium current sheet, the system commences to evolve.
Magnetic reconnection is initiated gradually from the perturbation region, which produces two reconnection outflows and forms an arcade of newly-reconnected magnetic loops anchored at the bottom boundary where the magnetic field has been set to be line-tied on the photosphere.

The governing resistive MHD equations are as following:
\begin{eqnarray}
\frac{\partial\rho}{\partial t} & + & \nabla\cdot(\rho \mathbf{v})=0, \label{eq:mass} \\
\frac{\partial\rho\mathbf{v}}{\partial t} & + & \nabla\cdot(\rho\mathbf{v}\mathbf{v}-\mathbf{B}\mathbf{B}+\mathbf{P^{*}})=0, \label{eq:momentum} \\
\frac{\partial \mathbf{B}}{\partial t} & - & \nabla\times( \mathbf{v}\times \mathbf{B})=\eta_{m}\nabla^{2} \mathbf{B}, \label{eq:induction} \\
\frac{\partial E}{\partial t} & + & \nabla\cdot[(E+P^{*})\mathbf{v} - \mathbf{B}(\mathbf{B}\cdot\mathbf{v})]=S, \label{eq:energy}
\end{eqnarray}
where ${\mathbf{P^{*}}}$ is a diagonal tensor with components $P^{*} = P + B^{2}/2$ (with P the gas pressure), and $E$ is the total energy density
\begin{equation}
E=\frac{P}{\gamma-1} + \frac{1}{2}\rho v^2 + \frac{B^2}{2}, 
\end{equation}
${\gamma = 5/3}$ is the adiabatic index,
and the energy source term $S = \mu_0 \eta_{m}{j^2} + \nabla_{||}\cdot\kappa\nabla_{||} T$, which includes Ohmic dissipation and thermal conduction. Here $\mu_0, \eta_m$ and $\kappa$ are the magnetic permeability of free space, magnetic diffusivity, and parallel component of Spitzer thermal conduction tensor.

We normalize above MHD equations to non-dimensional forms using characteristic values in our calculations: ${\mathbf{B}^* = \mathbf{B}/B_0}, {\rho^*=\rho/\rho_0}, {p^*=p/(B_0^2/\mu_0)}, {\mathbf{v}^*=\mathbf{v}/V_0}, {T^*=(\beta_0/2)(T/T_0)}$, and ${\mathbf{J}^*=\mathbf{J}/J_0}$ with ${V_0=B_0/\sqrt{\mu_0 \rho_0}}$ and ${\beta_0=2\mu_0 p_0/B_0^2}$. Here an ambient gas pressure $p_0$ depends on $\beta_0$ which will be assigned in the following parameter Table \ref{tab:mathmode}. We set ${L_0 = 7.5 \times 10^4}$ km, $B_0=0.004$ Tesla, $\rho_0=1.93\times 10^{-11}$ kg/m$^3$, $T_0=2$ MK, $t_0= 92.3$ s, $V_0= 812.4$ km/s, and $J_0 = 4.2 \times 10^{-5}$ Am$^{-2}$ to model a typical solar flare in following simulations.

\subsection{Code description}
We use the Athena code to numerically solve the above MHD equations. 
Athena is a publicly available grid-based code for astrophysical MHD (Stone et al. 2008). 
In this work, we include Ohmic resistivity in MHD equations, and omit gravity and optically thin cooling. The code is based on the directionally unsplit high order Godunov method, and combines the corner transport upwind (CTU) and constrained transport (CT) methods. It provides superior performance for capturing shocks as well as contact and rotational discontinuities. The low numerical diffusion feature of Athena code provides a significant advantage for resistive MHD simulations of processes like magnetic reconnection.


\subsection{Initial and boundary conditions}
\begin{figure}[h]
\plotone{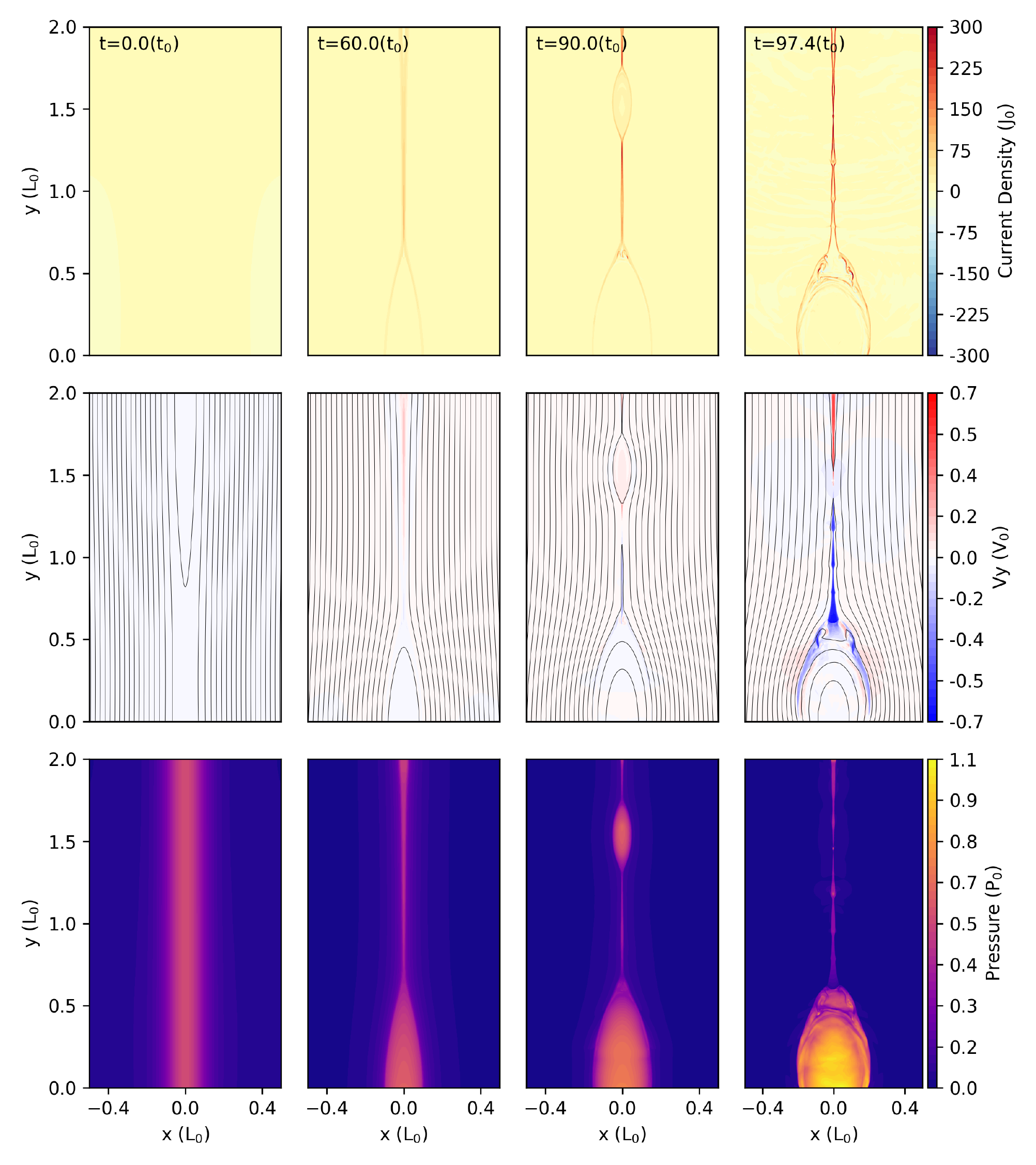}
\caption{The evolution of current-density magnitude $|J|$, $y$ component of velocity $V_y$ and pressure $P$. At the early phase ($t = 0, 60 t_0$), magnetic reconnection steadily takes place. After $t = 90$$t_0$, rapid reconnection appears as plasmoids develop inside the current sheet. Later, the reconnection becomes violent and more fine structures appear in this unsteady phase. 
\label{fig:whole_evol}}
\end{figure}

The initial configuration includes a pre-existing Harris-type current sheet along $y$-direction with the non-dimensional width $w=0.1$ as follows: 
\begin{eqnarray}
B_x(x, y) = 0,\\
B_y(x, y) = tanh(x/w), \\
p(x, y) = (1+\beta_0-B_y(x,y)^2)/2, \\
\rho(x, y) = 2p(x, y)/\beta_0.
\end{eqnarray}
In order to have the system evolve rapidly from the initial steady state to a bursty reconnection phase, we introduce perturbation magnetic field $B_{1x}$ and $B_{1y}$ on this pre-existing current sheet as follows:
\begin{eqnarray}
B_{1x}=\frac{2\pi}{L_y}A_{pert}cos(\frac{\pi x}{L_x})sin(\frac{2\pi (y-y_c)}{L_y})B_0, \label{eq:B1x}\\
B_{1y} = -\frac{\pi}{L_x}A_{pert}sin(\frac{\pi x}{L_x})cos(\frac{2\pi (y-y_c)}{L_y})B_0.
\label{eq:B1y}
\end{eqnarray}
Here $A_{pert} = 0.01$ is the non-dimensional perturbation strength, $y_c=0.25$ is the perturbation center along the $y-$ axis. $L_x$ and $L_y$ are non-dimensional perturbation wavelength which are set to 2 and 3.5 in order to minimize perturbations at boundaries. The initial current density, magnetic configuration and pressure distribution are plotted in Figure \ref{fig:whole_evol}. 




The boundary conditions are arranged in this following way. The right ($x=1.0$), left ($x=-1.0$), and top sides ($y=2.0$) of the simulation domain are open or free boundaries on which the plasma and the magnetic flux are allowed to enter or exit freely, 
and the boundary at the bottom ($y=0.0$) ensures that the magnetic field is line-tied,
\begin{equation}
\frac{\partial B_y(x, 0, t)}{\partial t} =0, \label{eq:line-tied}
\end{equation}
and the plasma does not slip and is fixed to the bottom boundary,
\begin{equation}
{\bf v}(x, 0, t) = 0, \label{eq:no-slip}
\end{equation}
respectively. For gas pressure and plasma density, we set
\begin{equation}
\frac{\partial P(x, y, t)}{\partial y} \mid _{y=0} = 0, 
\end{equation}
and
\begin{equation}
\frac{\partial \rho(x, y, t)}{\partial y} \mid _{y=0} = 0.
\end{equation}
We also set $J_{z}$ vanish at the bottom boundary as well, namely
\begin{eqnarray*}
J_z = (\frac{\partial B_y}{\partial x} - \frac{\partial B_x}{\partial y}) \mid_{y=0} = 0 \hspace{5mm} \mbox{for} \hspace{5mm} t > 0.
\end{eqnarray*}
Then $B_x(x, 0, t)$ can be set according to the above condition.

\subsection{Parameter table}
The parameters of our numerical simulation cases are listed in Table 1. 
In Cases 1 and 2, we fix the ambient magnetic field strength to $B_0$ and set different gas pressure to change the background plasma $\beta_0 = p_0/({B_0}^2/2\mu_0)$.
In Cases 3, 4, and 5, we introduce a uniform $B_z$ in the whole calculation domain to study the effect of magnetic guide field on TSs. In Case 6, we include anisotropic thermal conduction to compare with Case 7. 
We also perform simulations from different initial current sheets in Cases $6\sim 7$ by setting a sine-type \footnote{${B_y(x,y)=sin(\pi x/2w)}$ when $|x| <=w$, and ${B_y(x,y)=sign(x)}$ when $|x|>w$ \citep[see][in detail]{Shen2011}.} current sheet. Driven by same perturbations, these narrow sine-type sheets quickly evolve to the fast-reconnection phase when the length-width ratio of the current sheet exceeds a critical value (e.g.,\cite{Ni2010}) to allow plasmoids to appear. In the late phases, similar evolution features including reconnection outflows and plasmoids are found in both types of initial configuration. This indicates that the dynamical TS naturally forms in classic flare configurations and is insensitive to the particular initial setting.
In all seven cases, we use a uniform resistivity $\eta_{m}$ on the whole computational domain, which gives a constant magnetic Reynolds number $R_m = 10^5$.

\begin{deluxetable*}{ccCccc}[b!]
\tablecaption{Parameters and setup for all simulation cases\label{tab:mathmode}}
\tablecolumns{6}
\tablenum{1}
\tablewidth{0pt}
\tablehead{
\colhead{Cases} &
\colhead{Grids} &
\colhead{Background $\beta_0$} &
\colhead{Constant B$_z$} & 
\colhead{Thermal Conduction} \\
\colhead{} & \colhead{}
}
\startdata
1 & 2048 2048 & 0.1 & 0 & No & \\
2 & 2000 2000 & 0.02 & 0 & No & \\
3 & 2048 2048 & 0.1 & 0.05 & No & \\
4 & 2048 2048 & 0.1 & 0.25 & No & \\
5 & 2048 2048 & 0.1 & 0.5 & No & \\
6 & 2048 2048 & 0.1 & 0 & Yes & \\
7 & 2048 2048 & 0.1 & 0 & No \\
\enddata
\tablecomments{Magnetic Reynolds number $R_m=10^5$. Cases 6 $\sim $ 7 are initialized from sine type current sheets (see details in \cite{Shen2011}).}
\end{deluxetable*}

\section{Results}
Driven by the initial perturbation on magnetic fields defined in Equations (\ref{eq:B1x}) and (\ref{eq:B1y}), 
the initial current sheet becomes thinner and more intense, and eventually develops magnetic islands. In the following analysis, we focus on Case 1, and show the common features of dynamical termination shocks. In section 3.1-3.3 we present detailed analysis on the properties and dynamic features of the termination shock. In section 3.4, we will discuss effects of magnetic guide field, and present simulations including classical Spitzer thermal conduction \citep{Spitzer1962} in section 3.5. 

Figure \ref{fig:whole_evol} shows the evolution of the magnitude of current density $|J|$, velocity in the $y$-direction $V_y$, and pressure $P$. 
In the beginning, the magnetic field starts to diffuse at the X-point where perturbation is introduced. The magnetic fields at two sides of the current sheet slowly move toward one another due to the Lorentz force attraction between the field lines of opposite polarity. A pair of magnetic outflows gradually form and closed magnetic loops accumulate below the reconnection site due to the reconnected magnetic flux. This process gradually squeezes the initial current sheet, and the sheet narrows until it becomes intensified ($|J|$ is higher than 300 $J_0$ after time 90 $t_0$) and thin enough that the tearing mode grows and becomes nonlinear (e.g., see Furth et al. 1963; Loureiro et al. 2007; Ni et al. 2010). At this time, many magnetic islands appear inside the sheet and reconnection suddenly becomes violent. The reconnection outflow
speed is significantly enhanced, reaching a maximum speed of 0.7 $V_0$.
The fast reconnection results in the rapid growth of closed magnetic loops, which would be observed as hot flaring loops (see Figure \ref{fig:whole_evol}). 

\begin{figure}[t]
\centering
\includegraphics[width=0.5\textwidth]{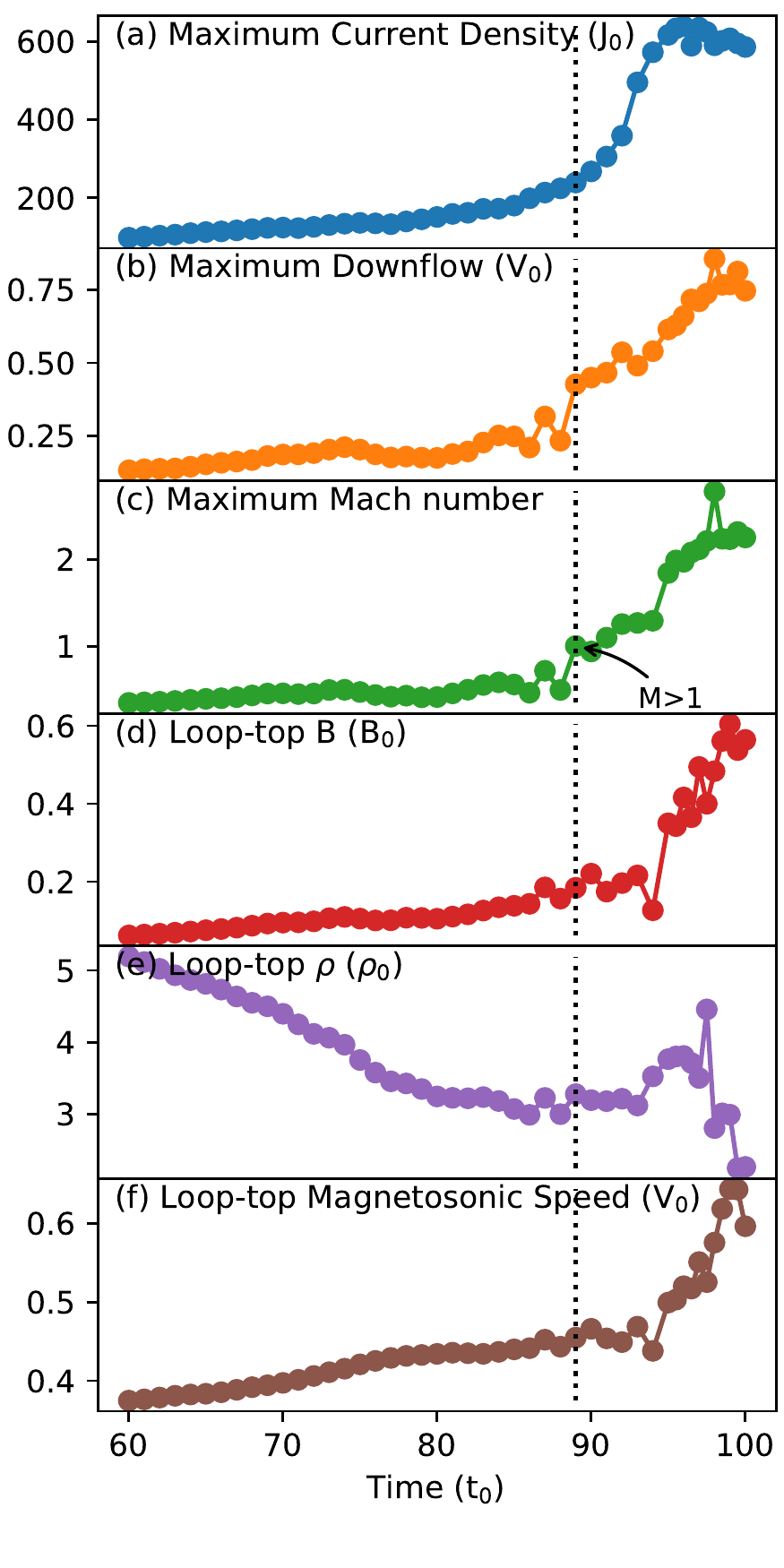}
\caption{The maximum current-density, downward outflow speed and
Mach number ($M = v/{(C_s^2+V_a^2)}^{1/2}$)
and average magnetic field strength, average density and average fast magnetosonic speed in loop-top regions within 0.1$L_0$ heights below the end point of downward outflows.
Here $v$ is plasma flow speed, $C_s$ is the local sound speed, and $V_a$ is Alfv\'en speed.
The dashed black line indicates the time ${89t_0}$ when the downward flow first exceeds the local fast magnetosonic speed. 
\label{fig:jzvyma_time}}
\end{figure}

We find that the TS develops when the downward plasma flow has a speed exceeding the local maximum fast-magnetosonic speed and it is stopped by the magnetic loop. As shown in Figure \ref{fig:jzvyma_time}, the reconnection downward outflow rapidly increases after time $89t_0$, and the outflow speed first exceeds the local fast magnetosonic speed along the reconnecting current sheet. During this period, the significantly enhanced current density indicates the reconnection becomes more efficient and violent as well. We also monitor properties in flare loop-top regions and show how average magnetic field strength, average density and average fast magnetosonic speed change with time in panels $(d)-(f)$. It is clear that amount of magnetic flux and density has been accumulated as the rapid reconnection was taking place, and a strong magnetic loop has been formed after $89t_0$ and becomes more dense after $94t_0$. After $95t_0$, the loop-top density decreases as the inflow plasma gets thinner during later phases of the fast magnetic reconnection. This relatively steady magnetic loop structure is important for the formation of TSs, as it becomes an obstacle that stops the downward super-magnetosonic outflows. Note we do not include chromospheric evaporation in the current study, which would greatly enhance the density of the post-flare loops and possibly, facilitate the shock formation as the local Alfv\'en speed is inversely proportional to $\sqrt{\rho}$.

\subsection{Jump condition and Morphology}
\begin{figure}[t]
\plotone{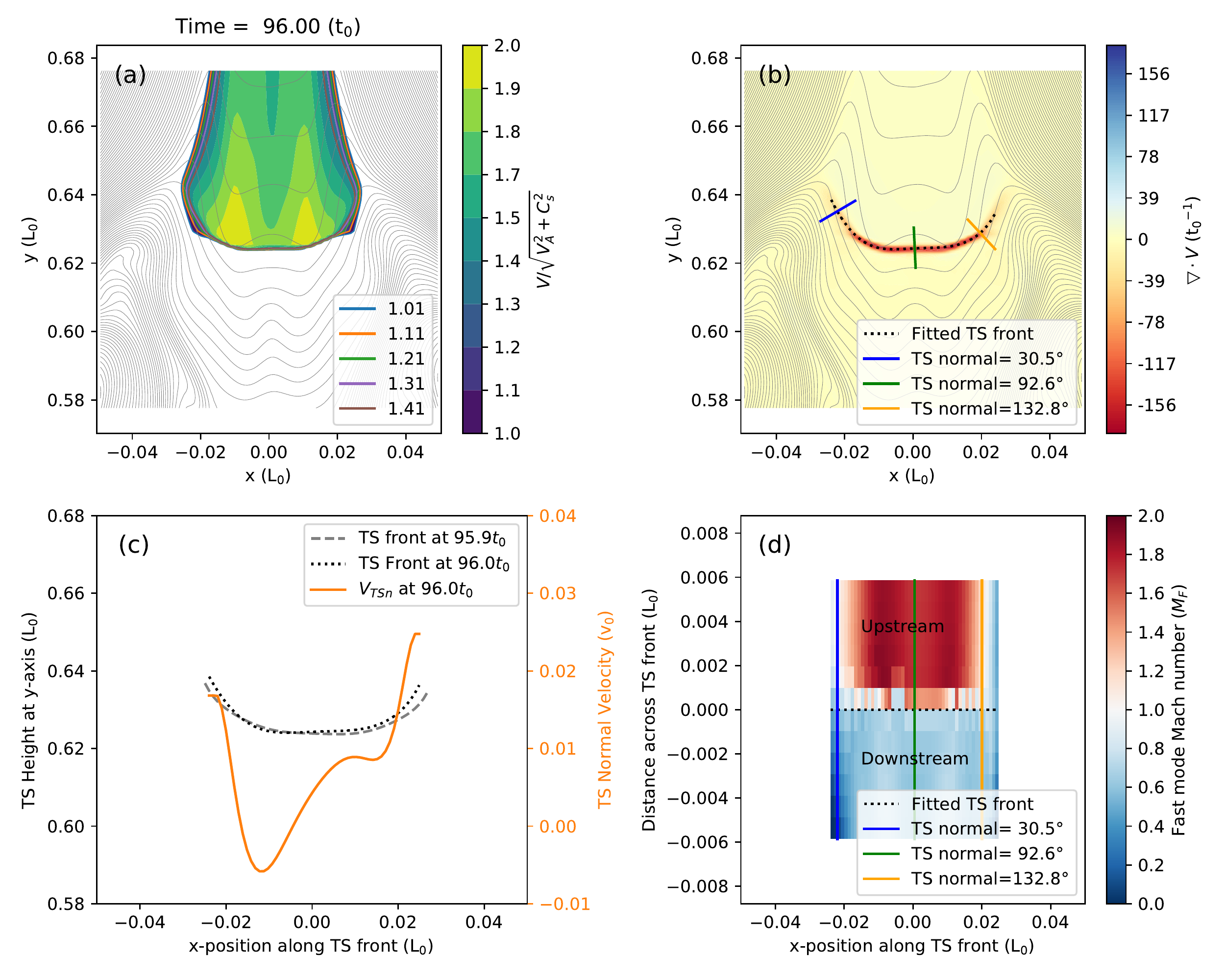}
\caption{
(a) Distribution of Mach number $v/{(C_s^2+V_a^2)}^{1/2}$, (b) Velocity divergence $\nabla \cdot v$ at time $t=96 t_{0}$. The black contour lines in panels (a) and (b) show magnetic field lines, and the red curve with the minimum $\nabla \cdot v$ indicates the position of TS front in panel (b). Three straight cuts (blue, green and orange lines) show the local TS normal directions in panel (b).  Panel (c) shows TS heights at times 95.9 and 96.0$t_0$, and the normal velocity $V_{TSn}$ at $t=96t_0$. Panel (d) shows the results across and along the TS at $t=96t_0$ . The horizontal axis and three sampling cuts (blue, green and orange lines) are same as in panel (b). \label{fig:divv_2d} }
\end{figure}

We first analyze the velocity distribution above the flare magnetic loops to identify the exact location of the TS front. The plasma velocity should be super-fast-magnetosonic on the upstream side and become sub-fast-magnetosonic on the post-shock (or downstream) side, respectively. 
Therefore, we plot the Mach number defined by $v/\sqrt{C_s^2+V_a^2}$, the ratio of plasma velocity over the maximum fast-magnetosonic speed perpendicular to $\mathbf{B}$ in Figure \ref{fig:divv_2d}(a) to show the surface of TS front.
It is clear that the downward reconnection flow Mach number steps from above 1.4 to less than 1.0 crossing a sharp boundary, which is the TS front. The TS front can also be easily identified by examining the velocity divergence (${\nabla \cdot v}$) map (Figure \ref{fig:divv_2d}(b)): it appears as a sharp layer with negative values of ${\nabla \cdot v}$, as strong compression is expected in the immediate vicinity of the shock.
In Figure \ref{fig:divv_2d}(b), the shock front is delineated in the ${\nabla \cdot v}$ map as a thin, red curve, and its location matches well with the sharp discontinuity in the Mach number map in Figure \ref{fig:divv_2d}(a).

We then represent primary variables in the co-moving frame with the TS front and compute the fast mode Mach number ($M_F$). Based on the ${\nabla \cdot v}$ map, we can fit the TS front profile using 8 order polynomial fitting as shown by dotted line in Figure \ref{fig:divv_2d}(b), and obtain the spatial position of the TS front. 
We then estimate the TS normal velocity (${v_{TS}}_n$) using backward difference in time between current and previous frames. For each point on the TS front, we find its previous position (e.g., $t=95.9t_0$) along the TS normal direction, measure moving distance of the chosen point during this period, and obtain ${v_{TS}}_n$ (see Figure \ref{fig:divv_2d}(c)).   
At each position along the TS front curve, we choose a set of sampling cuts along the local TS normal direction as shown in Figure \ref{fig:divv_2d}(b), and obtain the normal component of primary variables along these cutting lines.
Then the fast mode Mach number can be calculated by: $M_F = |{v_{TS}}_n - v_n|/C_F$. Here ${v_{TS}}_n$ is the normal TS velocity at each sampling cut, $v_n$ is the normal velocity of plasma flows to the TS, and $C_F = [\frac{1}{2}(C_s^2 + V_a^2 + \sqrt{{(C_s^2+V_a^2)}^2 - 4C_s^2 V_a^2 cos \theta_{Bn} ^2}]^{1/2}$ is the local magnetosonic fast-mode wave speed, and $\theta_{Bn}$ is the angle between the local magnetic field and the shock normal. 
We show distributions of $M_F$ along these chosen sampling cutting lines in Figure \ref{fig:divv_2d}(d). Here the horizontal axis is for $x-$ location of the TS, and the perpendicular axis is the distance away from the TS front along each of cutting lines. It can be seen that $M_F$ sharply jumps between upstream and downstream regions. For example, along the green cutting line, $M_F$ is as high as $\sim$ 1.8 on the upstream side and less than one on the downstream side. This jump sharply occurs in $2 \sim 3 $ grid sizes. At the left and right ends along the TS front, the compression becomes very weak as shown on the ${\nabla \cdot v}$ map. Therefore, we can see the fast mode Mach number reduces to one at the blue and orange cuts, and even less than one outside these two lines. We then can estimate the expansion length of the TS front in $x-$ direction, which ranges from $\sim -0.023$ to $\sim 0.02$ in the condition $M_F > 1$. 

\begin{figure}[h]
\centering
\plotone{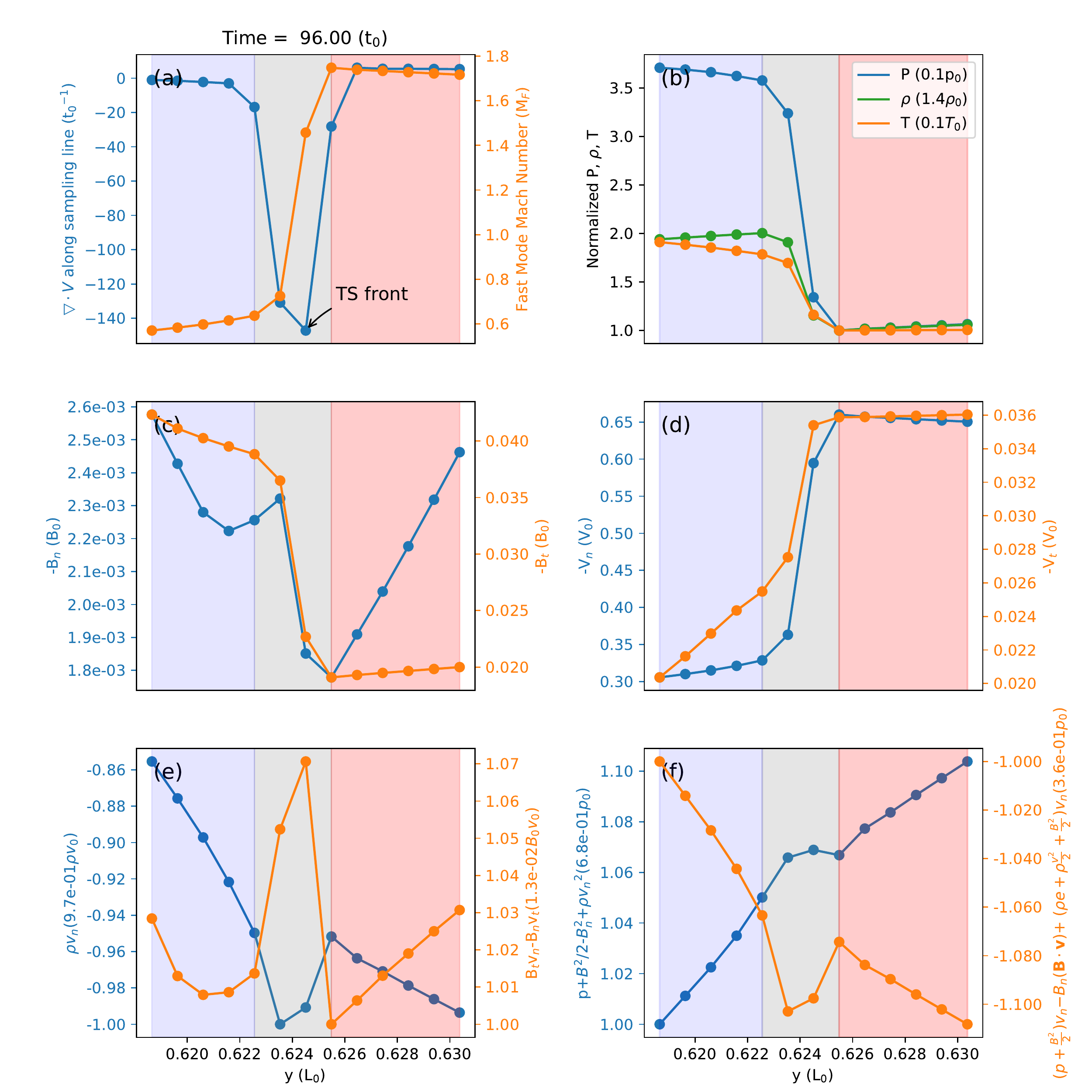}
\caption{Variation of primary variables across the shock front along the TS normal direction at $x=0$ in the co-moving frame: (a)${\nabla \cdot v}$ and $M_F$; (b)pressure $p$, density $\rho$ and temperature $T$; (c) magnetic field components $B_n$ and $B_t$; (d) velocity components $v_n$ and $v_t$; (e) mass flux and magnetic flux; and (f) momentum flux and energy flux. The red and blue shades indicate pre-shocked and post-shocked regions, respectively. The gray shade is for the TS front. \label{fig:jumpcondition} }
\end{figure}

%
%

Furthermore, we investigate other MHD variables across the TS front. 
These quantities shown in Figure \ref{fig:jumpcondition} include:
(a) velocity divergence (${\nabla \cdot v}$) and fast-mode Mach number ($M_F$), (b) gas pressure ($p$), density($\rho$) and temperature($T$), (c) normal component $B_n$ and transverse component $B_t$ of magnetic field, (d) velocity components $v_n$ and $v_t$, (e) mass flux component $\rho v_n$ and magnetic flux component $B_t v_n - B_n v_t$,  
(f) momentum flux $p+B^2/2-{B_n}^2+\rho v_n^2$ and energy flux $(p+B^2/2)v_n - B_n(\mathbf{B} \cdot \mathbf{v}) + (\rho e + \rho v^2/2 + B^2/2)v_n$.

The transition of pressure between upstream and downstream regions is very sharp, and jumps from 0.1 to around 0.38. The location of the shock front also can be identified by the minimum ${\nabla \cdot v}$, which matches well with the pressure jump. Similar sharp jumps can also be found in density, temperature, and the transverse magnetic component $B_t$ and normal velocity competent $V_n$ profiles in Figure \ref{fig:jumpcondition}(b) $\sim$ (d). It is worth noting that we do not resolve the real dissipation scale of the shock front (e.g., the order of ion skin depth) in these MHD numerical experiments because that scale is extremely narrow in high temperature plasmas such as the solar corona. Nevertheless, the transition still can be well-approximated as a discontinuous change between two sides of the shock front. As we can observe in Figure \ref{fig:jumpcondition}(a), the discontinuity is not infinite thin, with a thickness of roughly $2 \sim 3$ cells, which is a reasonable value in the shock capture MHD scheme. In Figure \ref{fig:jumpcondition}, we show this area with 3 cells as gray shades.
%
%
%
%

It is noticed that the magnetic field component $B_n$ (blue curve in Figure \ref{fig:jumpcondition}(c)) is small relative to the transverse component $B_t$, 
but it does not vanish either on the upstream or downstream sides. A horizontal flow $v_t$, albeit very small, can also be clearly seen (orange curve in Figure \ref{fig:jumpcondition}(d)). That indicates the TS is nearly perpendicular in the particular position we picked (also see green line in Figure \ref{fig:divv_2d}(b)), and the shock is dynamically evolving with weak horizontal flows. 

Along the cut, the momentum component and magnetic flux component are roughly constant between upstream and downstream regions. As can be seen in Figure \ref{fig:jumpcondition}(e), 
the relative changes for $\rho v_n$ and $B_t V_n - B_n V_t$ are roughly $1\%$ and less than $\sim 2\%$ between the upper edge and lower edge of the TS region (gray shaded regions in Figure \ref{fig:jumpcondition}), respectively. The momentum flux and energy flux parallel components are also conserved,
and the relative changes between upstream and downstream are still less than $\sim 2\%$ as shown in panel (f).
%
%
%
%
It is noticed that there are unavoidable measurement errors when estimating TS raising velocity and computing TS normal direction, which causes slight deviations from the classic prediction of a steady fast-mode shock. However, conservation conditions of primary variables are satisfied between upstream and downstream.

%
%

The shape of the termination shock is highly variable in time during this unsteady magnetic reconnection phase. As the plasmoids emerge and develop, reconnection becomes intermittent (see also \citep{Shen2011}) and the speed of the corresponding reconnection outflow changes with time.   
%
%
In addition, the development of plasmoids inside the current sheet has significant impact on both the outflow itself and the loop top structure.
The downward outflows tend to be separated by growing plasmoids into multiple fragment with different velocities. Once a plasmoid collides with the closed magnetic loop and merges into previous loops, the magnetic structure of loop is strongly perturbed. We found that both temporal outflow speeds and complex loop-top structures could change the shape of TS front.

\begin{figure}[h]
\plotone{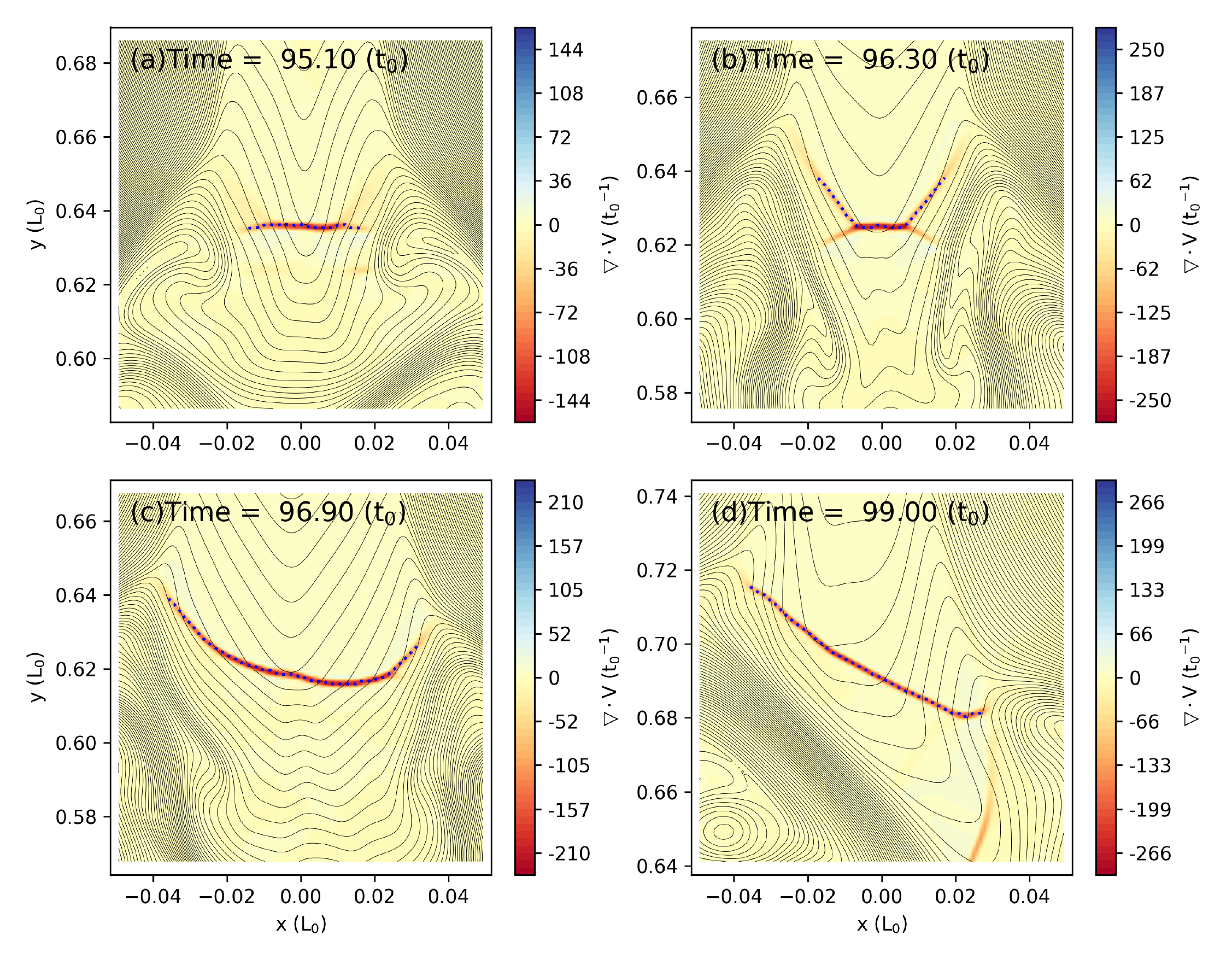}
\caption{Divergence of fluid velocity at four different time frames showing the shapes of TS front. The dotted lines indicate the edge of the shock front where primary variables jump crossing shock front. The TS shape is strongly modified by the intermittent reconnection outflow and loop-top structure. \label{fig:ts_geom}}
\end{figure}

In the symmetric case for a classic flare model, it can be expected that TS is a perfectly perpendicular shock at the symmetry center (or the $y-$axis in this paper) where the magnetic field only has a horizontal component. In other words, the magnetic field direction both in the upstream and downstream sides is parallel to the TS front at the symmetry center. 
However, the magnetic configuration could be asymmetric in many realistic environments, which may cause variations on the TS geometry. Instead of investigating the asymmetric configuration itself, we show here variable TS slopes when the system evolves to the asymmetrical case due to the numerical perturbations.
In Figure \ref{fig:ts_geom}, we display the shape of the TS illustrated by $\nabla \cdot v$ at four different times. The shock fronts show time-varying shapes including quasi-flat, horizontal and oblique components, curves, and oblique-flat shapes. In panel (b), a pair of sub-shocks can be seen below the horizontal TS front at time 96.30$t_0$. It could appear as the reflection of plasma flows becomes strong below the TS, and disappear once there are not reflected flows in the downstream regions.
In the following analysis, we focus on the upper shock structure which is directly driven by the reconnection downflow, if multiple sub-shock structures are present at the same $x$ coordinate.
%
%

\subsection{Dynamics of TSs}



\begin{figure}[h]
\centering
\includegraphics[width=0.5\textwidth]{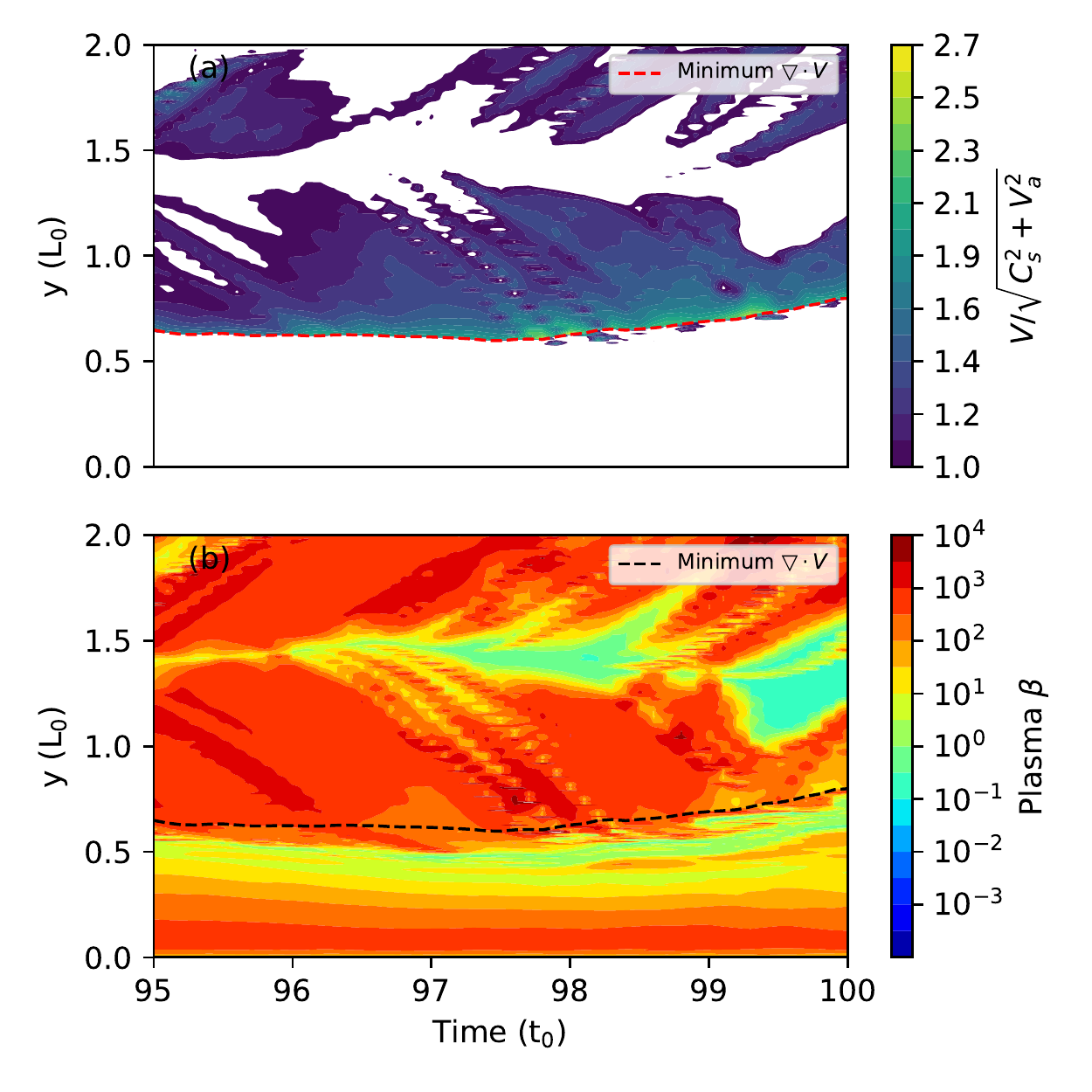}
\caption{Distributions of (a) flow Mach number $v/\sqrt{C_s^2+V_a^2}$ and (b) plasma $\beta$ along the y-axis ($x=0$) as functions of time. The dashed lines show minimum $\nabla \cdot v$, which indicate the height of TS fronts.
%
%
\label{fig:ma_beta_time}}
\end{figure}

In Figure \ref{fig:ma_beta_time}(a), we plot the distribution of the flow Mach number ($ v/\sqrt{C_s^2+V_a^2} > 1$) along the $y$ axis ($x=0$), which is the direction of the reconnection current sheet, as a function of time between $t=95t_{0}$ and $100t_{0}$ ($\sim$7.5 minutes duration)  --- a representation commonly referred to as a time-distance map or ``stack plot''. 
The dashed lines indicate the locations of the TS where $\nabla \cdot v$ is minimum.
During this period, the reconnection outflows are accelerated away 
from the X-point, and become super-magnetosonic in the exhaust region until they hit the top of the closed flaring loops. The acceleration of upward and downward moving plasmoids away from the X-point nearby $y=1.4$ is clearly seen.
It is clear that the edge of $v/\sqrt{C_s^2+V_a^2} >1$ separates the upstream side
and the downstream side of the TS around the height $0.6<y<0.75$. On the upstream side of the shock, the Mach number is frequently enhanced during this turbulent reconnection phase, and is directly affected by the downward outflows with higher velocity. 
%
%

The local plasma $\beta$ close to the TS appears to vary with time. The local plasma $\beta$ is clearly close to the values in magnetic reconnection downflows, and significantly larger than the plasma $\beta$ inside flare loops. As shown in Figure \ref{fig:ma_beta_time}(b), the lower range of the local plasma $\beta$ at the TS is around 100.
This confirms that the TS forms above magnetic loops and the shock properties are not strongly affected by the plasma inside magnetic loops.  
%
%

\begin{figure}[h]
\plotone{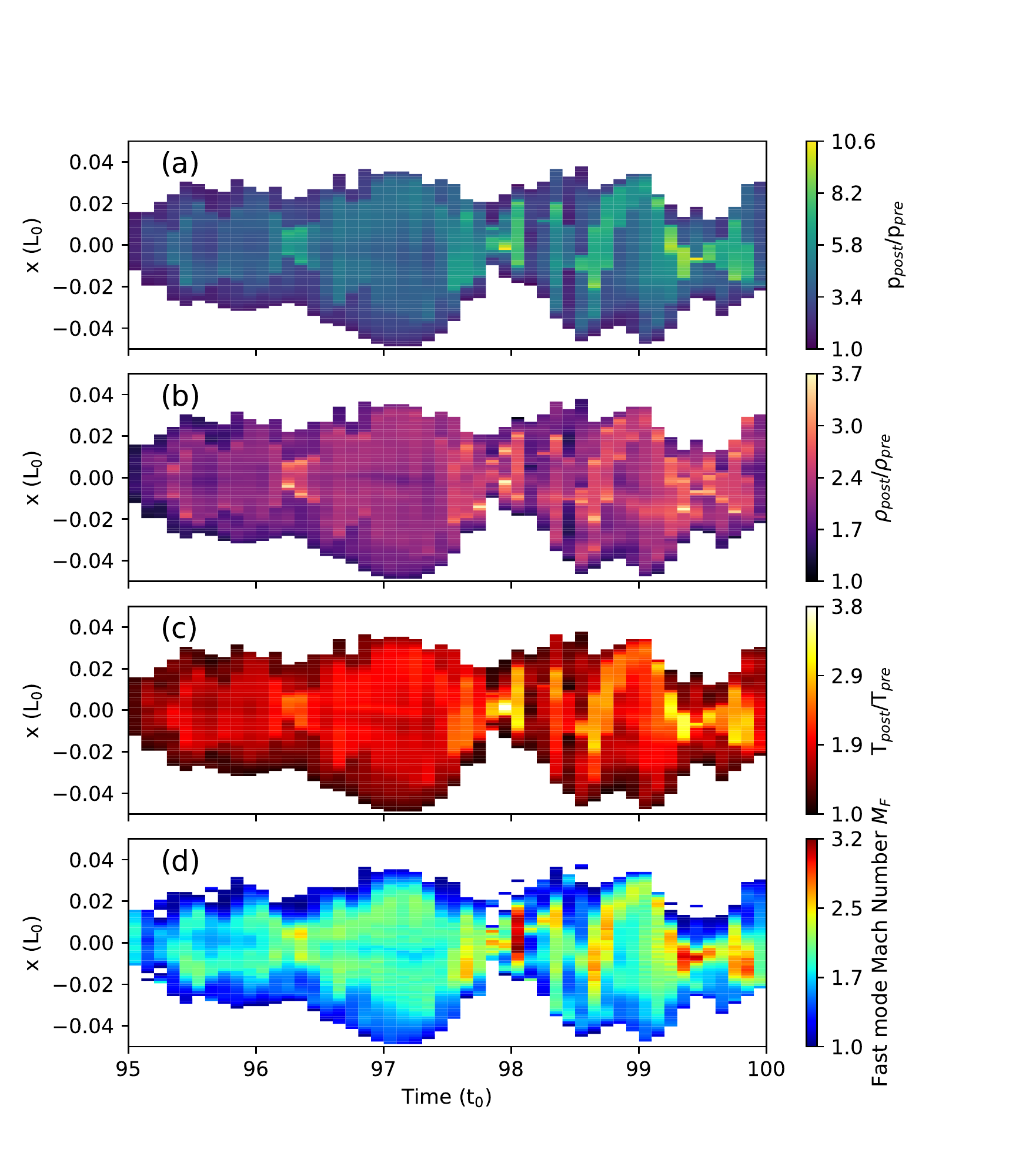}
\caption{
The compression ratio distribution across the TS front at different times. (a) Gas pressure ratio between post-shock and pre-shock sides; (b) and (c) are density ratio and temperature ratio; (d) is the fast mode Mach number $M_F$. The threshold condition $M_F > 1.0$ is used to set the plotting range along the TS front in this figure.
\label{fig:ratio_2d}}
\end{figure}

In addition, we calculate the compression ratio between upstream and downstream, and show here how the compression ratio changes spatially and temporally. In Figure \ref{fig:ratio_2d}, 
we plot gas pressure ratio, density ratio,  temperature ratio, and $M_F$ along the shock front.
We trace the TS front according to the minimum $\nabla \cdot v$ at each time, and obtain a set of sampling points at the TS front. For each sampling point, we then calculate primary variables on upstream/downstream sides by 2D spatial interpolation along the normal direction of TS front at this point. In Figure \ref{fig:ratio_2d}, the vertical axis shows the location of sampling points in the x-direction, and the horizontal axis is for time. During the plasmoid reconnection phase (95-100 $t_0$), the density compression ratio intermittently increases due to unsteady reconnection downward flows. The dominant density ratio ranges from 1 to 3 depending on time and location. The maximum density ratio can be as high as 3.7 near a corner of two oblique shock fronts where the downstream/upstream density could be slightly overestimated/underestimated due to interpolation process. The intermittent nature of the quantities for both spatial distribution and temporal distribution is clear. The variation of pressure and temperature ratios are also clear, and it is also easy to see that maximum value can larger than 10 and 3, respectively. 
The distribution of fast mode Mach number $M_F$ also varies in space and time, and ranges from $1 \sim 3$.

In a recent analytical CME/flare eruption model, \cite{Forbes2018} pointed out that the predicted fast mode Mach number is ${[2/(\gamma-1)]}^{1/2}$ \citep{Soward1982, Forbes1986a} for Petschek reconnection with an inflow plasma of zero $\beta$. In the case of $\gamma = 5/3$, this gives $M_F \approx 1.73$. In fact, the overall $M_F$ in our simulations agrees in general with their expected values as shown in Figure \ref{fig:ratio_2d}(d). Furthermore, $M_F$ in our simulation also can occasionally exceed the predicted value due to the occasion "bursts" of the outflow speed, which is associated with the development of tearing (or plasmoid) instabilities.

%
%
%
%

Along the shock front, the density compression dramatically varies with time. The spatial location of the maximum density compression ratio is not always at the point directly below the current sheet (i.e., $x=0$), but varies in time from $x=-0.02L_0$ to $x=+0.025L_0$. The highly variable nature of the density compression along the shock front is associated with the slope of the TS front, the bursty reconnecting downward flows, and/or the detailed properties of plasmoids. 
%
%
%
%
Another interesting feature is that the spatial position where maximum temperature ratio appears to depart from either the position of maximum pressure ratio or density ratio at some particular time. 
For example, the maximum temperature jump appears at around $x>0$ at $97.9t_0$, while the maximum density compression appears on another side $x<0$ and the maximum pressure ratio can not be found nearby these region ($x \sim 0$) as well.
This is not surprising since the magnetic field configuration and corresponding magnetic pressure are highly dynamical that causes the complex compression in downstream region than a steady TS scenario. 

\begin{figure}[h]
\centering
\plotone{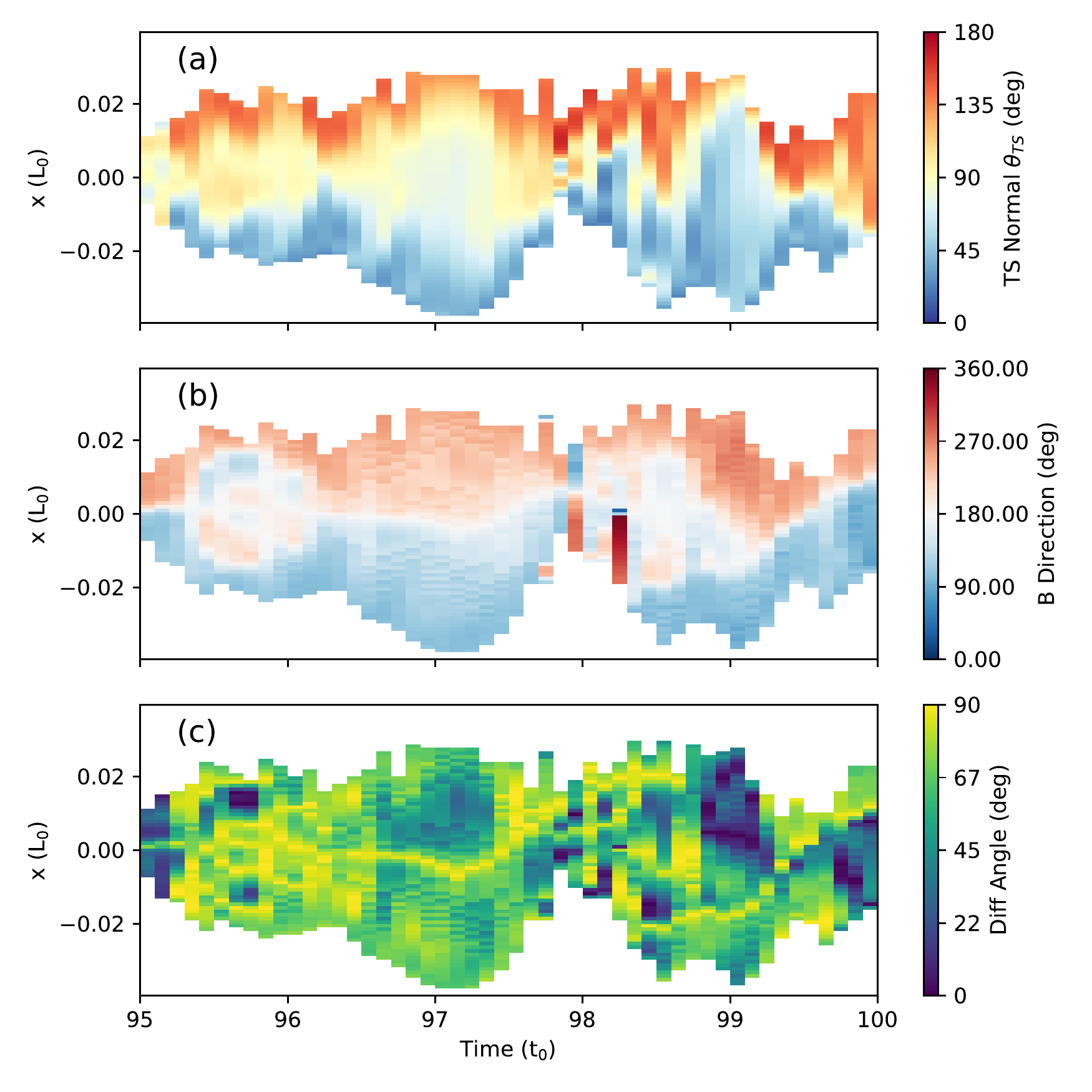}
\caption{(a) Normal direction of the TS front at different times. The angles are measured from the $x-$ axis direction to the normal direction; (b) Magnetic field direction at the shock front; (c) Interior angle between normal direction and magnetic field direction. The TS front is defined as in Figure \ref{fig:ratio_2d}.
\label{fig:normal_direction}}
\end{figure}

In Figure \ref{fig:normal_direction}, we display the normal direction of the shock front and examine the shock property during its dynamical evolution.
%
%
We measure the shock normal direction and magnetic field direction at each point along the shock surface, 
 and show them in Figure \ref{fig:normal_direction}(a) and (b). We plot the shock-normal angle, measured from the normal direction to the x-axis, and calculate the interior angle between TS normal direction and magnetic field direction in panel (c). 
It can be seen in panels (a) and (b) that the TS direction dynamically changes during time 95$t_0$ to 100$t_0$. At the system center ($x=0$), the shock front direction ranges from around 60 degrees to around 130 degrees. In panel (c), we can see that the interior angle is close to 90 degrees at the system center ($x=0$), where the TS can be regarded as a nearly perpendicular shock for most of the time. However, as shown in Figure \ref{fig:normal_direction}(c), it is by no means a stable perpendicular shock, as usually depicted in the standard flare cartoons (references to e.g., Shibata et al. 1995), even at $x=0$. Moreover, the TS becomes more and more oblique towards the edge of shocked surface, where the interior angle gradually decreases to $<30$ degrees.

\begin{figure}[h]
\centering
\includegraphics[width=0.6\textwidth]{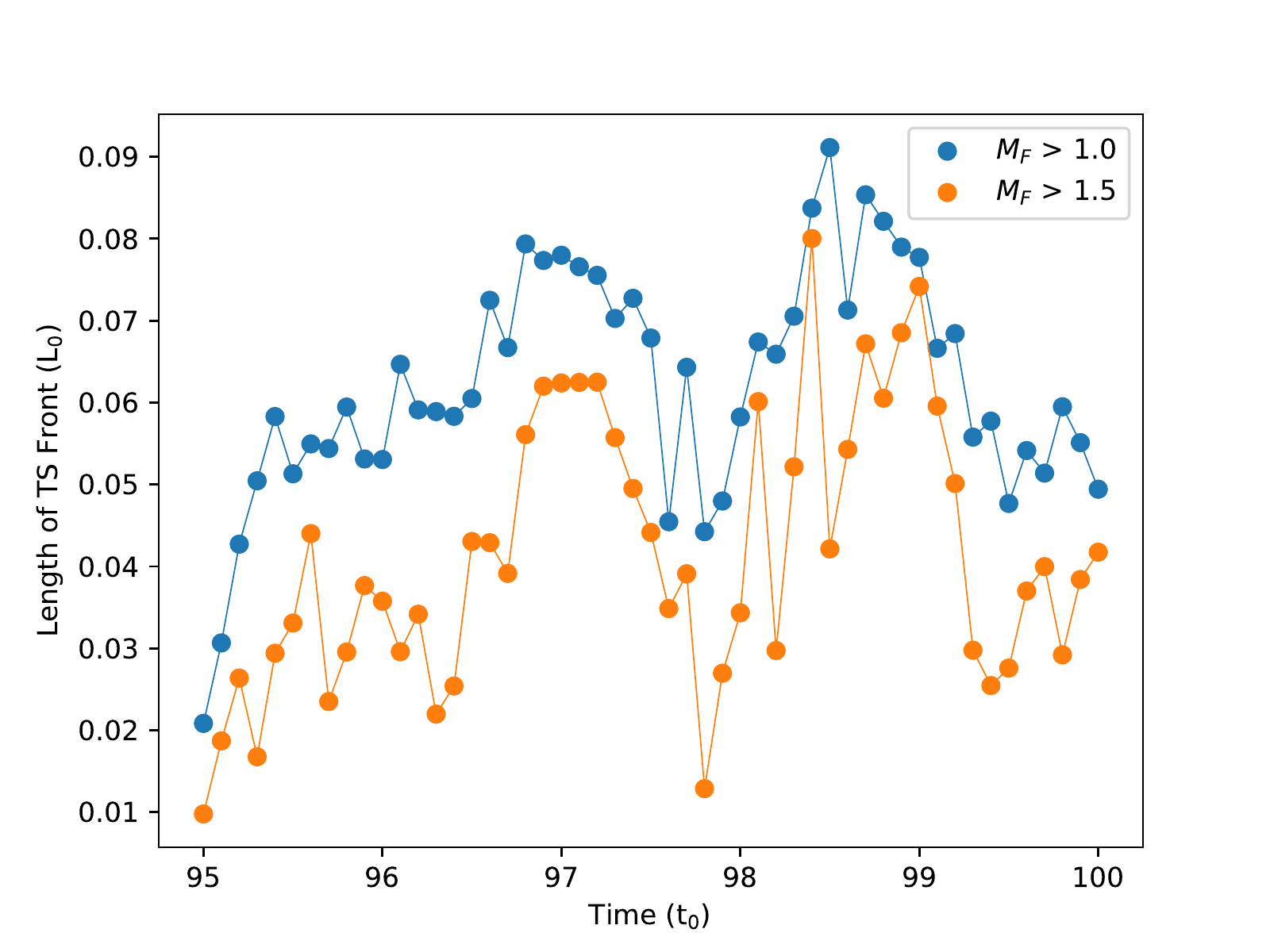}
\caption{The length of the TS front. The cyan line is for threshold of $M_F>1.0$, and orange line is for $M_F>1.5$. \label{fig:ts_length}}
\end{figure}

It is convenient to obtain the TS length measured along the TS front profile on the $\nabla \cdot v$ map. In Figure \ref{fig:ts_length}, we trace shock front profiles extending to the left and right sides from the system center ($x=0$), 
and define the end points according to the fast mode Mach number. The threshold value at the end points are chosen either as 1.0 or 1.5.
%
%
In Figure \ref{fig:ts_length}, the cyan line is for the threshold $M_F > 1.0$ and the orange line is plotted for higher threshold value $M_F > 1.5$. Because the evolution of the length is highly dynamical, the maximum length is larger than 0.09$L_0$ while the minimum length can be as short as around 0.01$L_0$ at a particular time. Scaling it according to the characteristic length ($L_0=7.5\times 10^4$ km) in Section 2, the TS length is in range of $<1$ Mm to $6.8$ Mm, which is consistent with the observed values in \cite{Chen2015} ($\sim 5$ Mm; c.f., their Fig. 3A).
%
%

\subsection{Effect of Plasmoids on Mach Number}
\begin{figure}[h]
\centering
\includegraphics[width=0.5\textwidth]{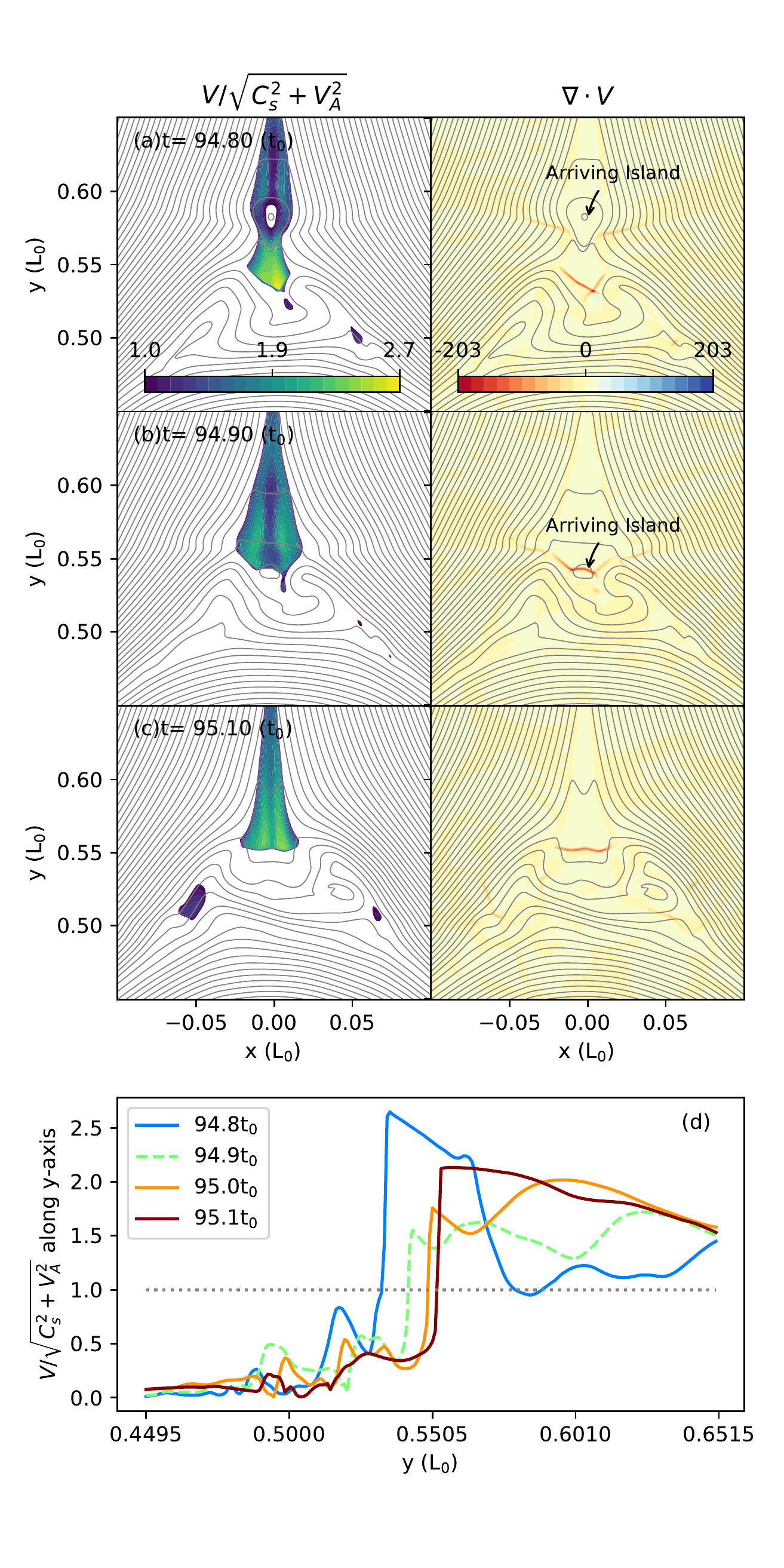}
\caption{Impacts of a small plasmoid on the termination shock.\label{fig:blob_ts_min}}
%
%
\end{figure}

\begin{figure}[h]
\plotone{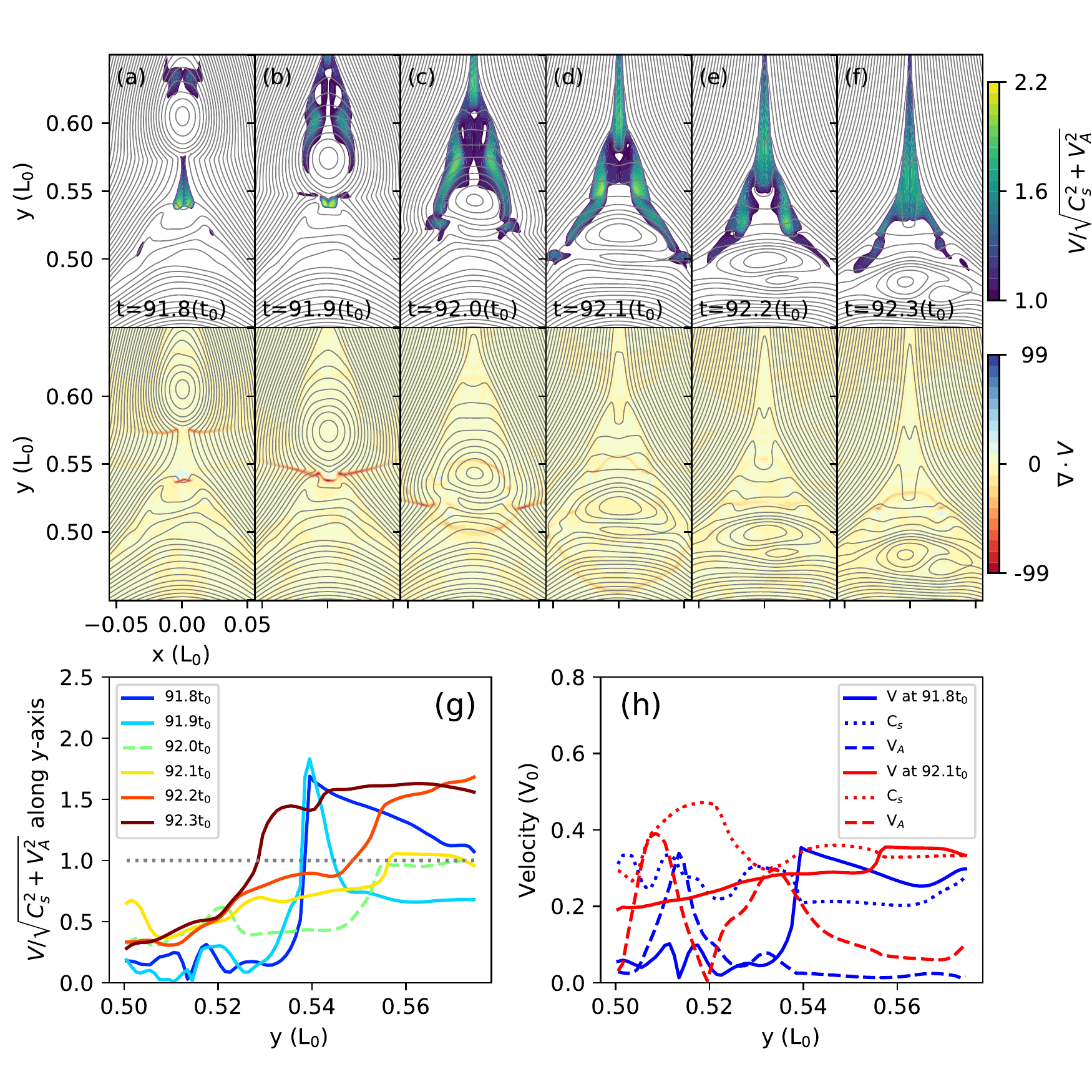}
\caption{The TS is destroyed by the well developed plasmoid, and then forms again above the plasmoid. Panels (a) $\sim$ (f) are Mach number and velocity divergence contours, and (g) shows Mach number distribution along the $y-$axis; Panel (h) shows absolute velocity, local sound speed and local Alfv\'en speed along the $y-$axis at two particular times using cyan lines(91.8$t_0$) and red lines(92.1$t_0$).
%
%
\label{fig:blob_ts_max}}
\end{figure}

As described above, we have shown that the collision between plasmoids and the closed magnetic loops can cause dynamic evolution of the TS. The shock front can temporally change its height and shape. In this section, we look in detail at the shock strength when the collision takes place. 
We choose two typical episodes in the following analysis from Case 2 in which the background magnetic field is relatively stronger due to the initial lower plasma $\beta$. In this case, we can see more small scale magnetic islands during the evolution, and choose two typical events to analyze the TS strength in detail.

In Figure \ref{fig:blob_ts_min}, we demonstrate how the Mach number of the TS changes when a small downward moving blob is merging into the closed magnetic loops. 
We color the downward flows with $v/\sqrt{C_s^2+V_A^2} > 1$ in panels (a)-(c), and plot the Mach number variations along the $y-$ axis in panel (d) for different times. At time 94.8$t_0$, the shock front is located at the height $y \sim 0.53L_0$ where the maximum Mach number is larger than 2.5. Meanwhile the downward moving magnetic island has arrived to height $y \sim 0.58L_0$. 
We can see a lower Mach number at the core of this magnetic island due to higher pressure and magnetic field than the ambient downward flows.
At time 94.9$t_0$, the center of this magnetic island arrives at the shock front. In the Mach number plot, it manifests itself as a slightly lower Mach number valley near the shock front. After this magnetic island has been totally merged into the previous closed magnetic loops, the TS front appears at a higher altitude, $y \sim 0.55L_0$. 
We can see more clearly the evolution history for this collision process in panel (d). In this panel, the dashed line is for time 94.9$t_0$ when the magnetic island is crossing the shock front. It is clear that the fast mode Mach number on the upstream side quickly decreases to around 1.5 from above 2.5. Once the magnetic island has merged into the magnetic loops, the Mach number rises again to 2.2 as shown by the dark green line as the shock moved up to a higher altitude.
%
%

Another collision event is shown in Figure \ref{fig:blob_ts_max}. In this case, the incoming plasmoid has fully developed for a relatively long time inside the current sheet and grows to a larger size. As this plasmoid accumulated magnetic flux and mass, it moved more slowly than the ambient outflow. In panels (a) and (b), we can see the plasma surrounding this magnetic island is sub-magnetosonic at times 91.8$t_0$ and 91.9$t_0$. At the same time, the TS front is at $y \sim 0.54$. Once this magnetic island reaches the shock front after time 92.0$t_0$, the TS front completely disappears, and the center region above the closed loops turns to sub-magnetosonic. In this collision process, the TS has been totally destroyed. At time 92.1$t_0$, a weak shock front re-appears at higher altitude above this magnetic island. At time 92.3$t_0$, we can see that a new TS has been restored as the plasmoid has merged fully into the underlying flare loops and the super-magnetosonic reconnection outflows build up in the upstream region again. The maximum Mach number then rises again to around 1.8. In panel (f), it is easy to see that the height of the new TS is $y=0.53$, which is even lower than the shock front before the collision. Therefore, the location of the TS front in $y-$direction can remarkable vary during a short period.
%
%
%
%

In this case, the variation of Mach number along the $y-$ axis is shown in 
Figure \ref{fig:blob_ts_max}(g). As the green dashed line shows, at time 92.0$t_0$ there is no TS front because the plasma is sub-magnetosonic. We can examine several characteristic speeds to explore why the shock front may disappear. 
In panel (h), we plot the plasma flow speed ($V$), sound speed($C_s$), and Alfv\'en speed ($V_A$) along $y-$ axis at two particular times, 91.8$t_0$ and 92.1$t_0$. At the early time, 91.8$t_0$, the shock front is at $y=0.538$ where the flow speed drops from 0.37 to around 0.1. Until time 92.1$t_0$, the magnetic island moves down to the height $y=0.531$. Because the magnetic island center has high gas pressure and magnetic field, the plasma flow speed is around 0.29$V_0$, lower than both the local Alfv\'en speed and sound speed. As shown by the red lines, the plasma speed is still significantly smaller than sound speed in surrounding regions.

\subsection{Termination Shocks with Guide Fields}
In this section, we compare the strength of the TS in Cases 3 through 5 with different strengths of the guide field. The following simulations are performed by adding an initially uniform guide field $B_z$.
The $B_z$ component might change strength during the evolution.
%
%
Therefore, $B_z$ could affect the local pressure equilibrium inside the current sheet significantly. The closed magnetic loops and corresponding local characteristic speed could, in turn, vary with the changing guide field.
In here, we consider three different initial values of the guide field $B_z = 0.05, 0.25$, and $0.5B_0$ (Cases 3, 4, 5 in Table 1) in the simulation zone. 
The corresponding background plasma $\beta_0$ is then equal to $p_{ini}/(0.5B_{ini}^2 + 0.5B_z^2)$, which are $\sim$ 0.0998, 0.094 and 0.08 in the initial setting of non-dimensional pressure $p_{ini}=0.05$ and non-dimensional magnetic strength $B_{ini}=1$. We then analyze the TS properties on the $y-$ axis for above three cases.

\begin{figure}[h]
\centering
\includegraphics[width=0.7\textwidth]{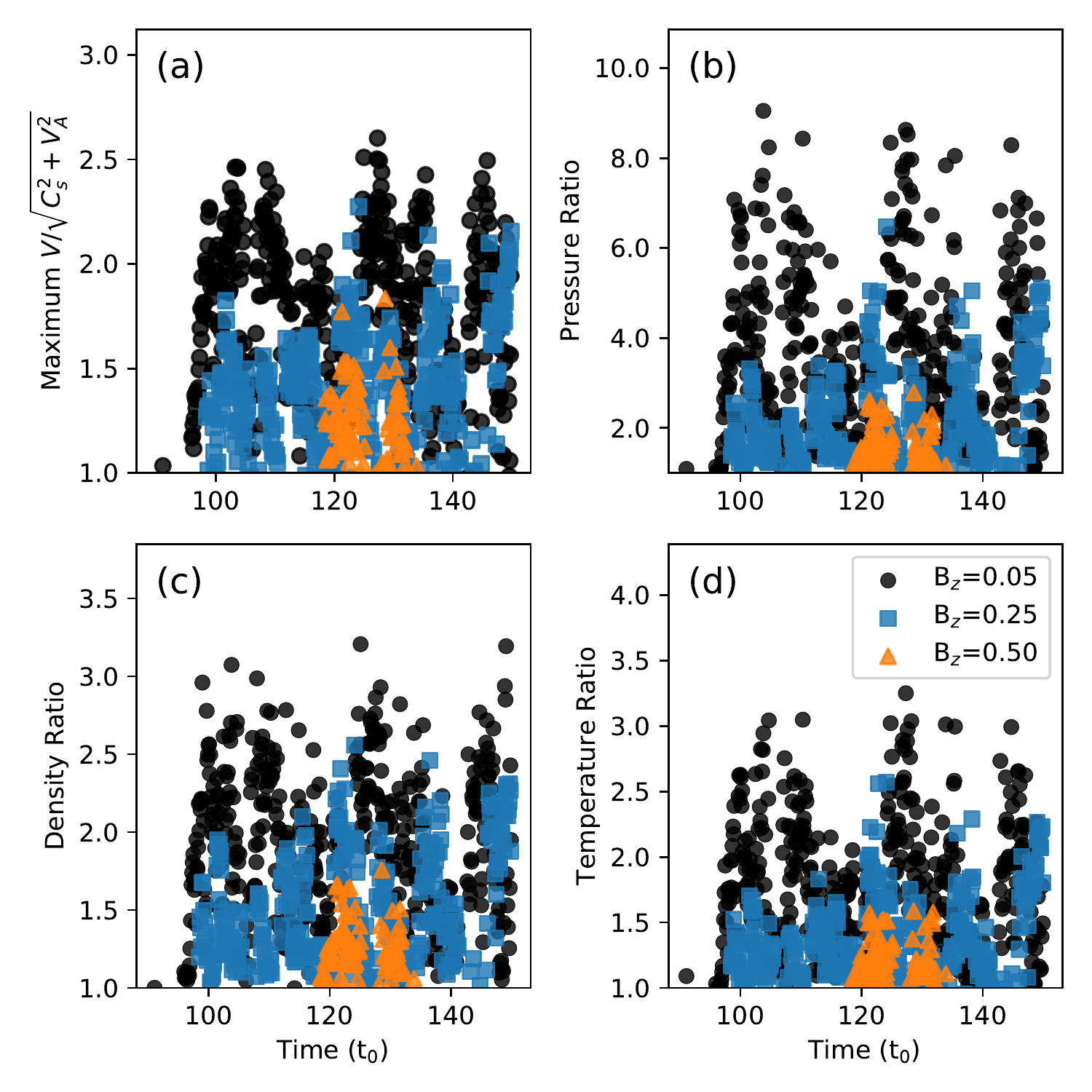}
\caption{(a) The maximum $v/\sqrt{C_s^2 + V_A^2}$ on the upstream side along the $y-$axis, and (b-d) compression ratios between upstream and downstream sides for these times with $v/\sqrt{C_s^2 + V_A^2} > 1$ along $y-$ axis. The initial guide fields are $B_z$ = 0.05, 0.25, and 0.5$B_0$ in the three cases (Cases 3-5 in Table 1). \label{fig:ma_bz}}
\end{figure}

Figure \ref{fig:ma_bz}(a) shows the maximum $v/\sqrt{C_s^2 + V_A^2}$ on the upstream side along $y-$ axis at different times. It is clear that the TS becomes weaker as the guide field increases to 0.5 (triangle points), and the corresponding maximum $v/\sqrt{C_s^2 + V_A^2}$ is less than $\sim 1.8$ throughout the time period.
In the strong guide field case ($B_z=0.5$), the upstream Mach number could be less than 1 for a long time. That means the TS more frequently disappears compared with the weak guide field cases ($B_z=0.05, 0.25$). 

Figures \ref{fig:ma_bz}(b)-(d) show compression ratios across the shock font in different guide field cases. The gas pressure ratio can be as high as 9 in low guide fields, while it only reaches around 3 in the strong guide case. The density and temperature ratios vary in range of 1.0 to 3.0 for the weak guide field cases and are less $\sim 1.8$ in the case with guide field $B_z = 0.5$.  

\subsection{Effects of Thermal Conduction}
In order to obtain a more realistic temperature and density distributions above the loop-top region, we introduce anisotropic thermal conduction in Case 6 to investigate the effects on the formation and dynamics of the TS. The evolution of the current sheet and its internal small-scale structures are different between the cases with and without thermal conduction. For example, small plasmoids appear at different times and positions in the two cases. We then choose a particular time when the TS appears at roughly the same height to compare the Mach number of the TS in Cases 7 and 6.

\begin{figure}[h]
\plotone{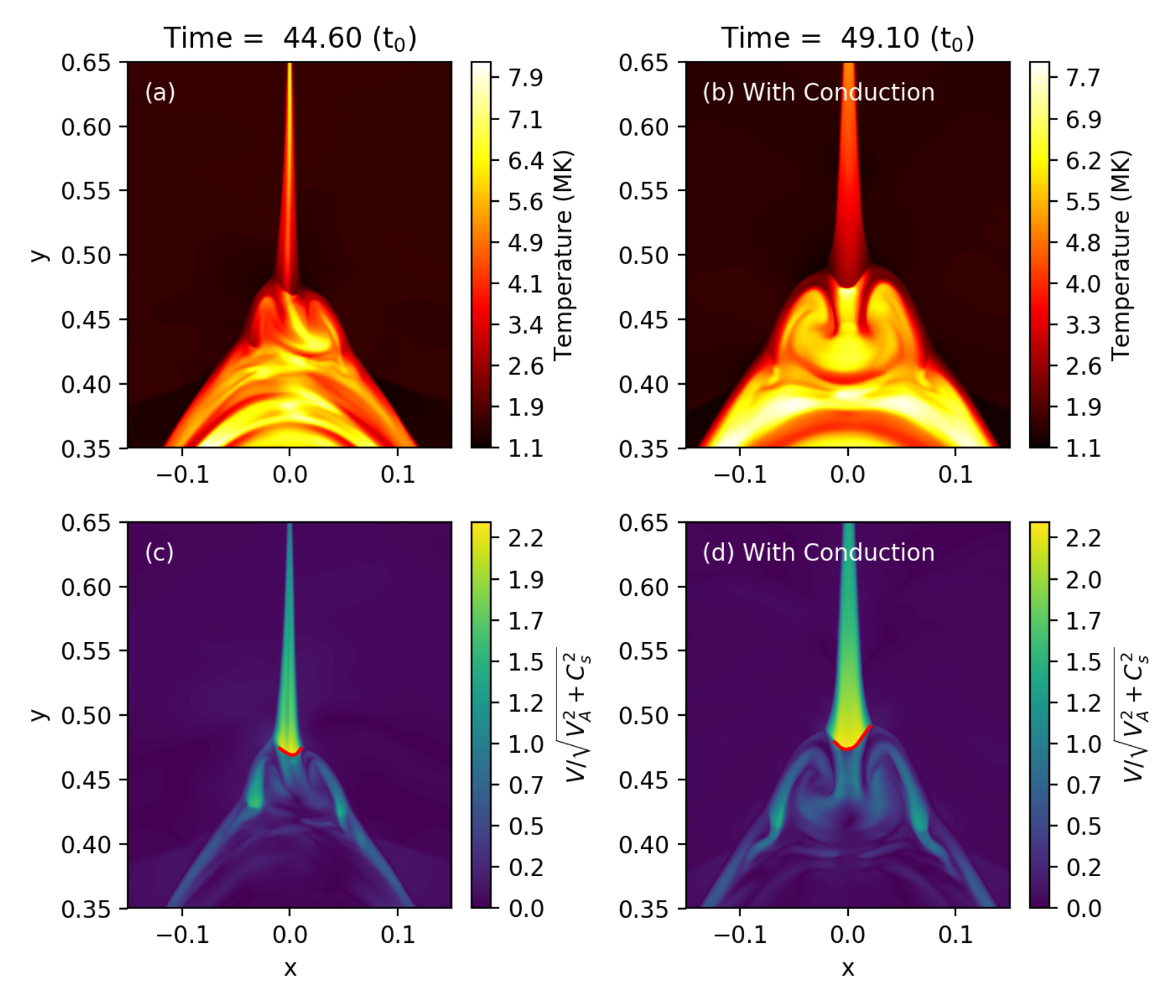}
\caption{Temperature and Mach number distribution in two simulations with (Case 6) and without (Case 7) thermal conduction. Red lines indicate TS fronts. \label{fig:conduction}}
\end{figure}

Figure \ref{fig:conduction} shows the temperature and Mach number distributions around the TS in these two cases. 
The Mach numbers are roughly in the same range $\sim 2.2$ at these particular times. The thermal conduction significantly changes the spatial distributions of temperature both in upstream and downstream regions. However, the maximum outflow speed above the TS is not strongly affected by the thermal conduction in the violent reconnection phases. It is then no surprise that the shock strengths are roughly the same in the two cases. 

\begin{figure}[h]
\centering
\includegraphics[width=0.5\textwidth]{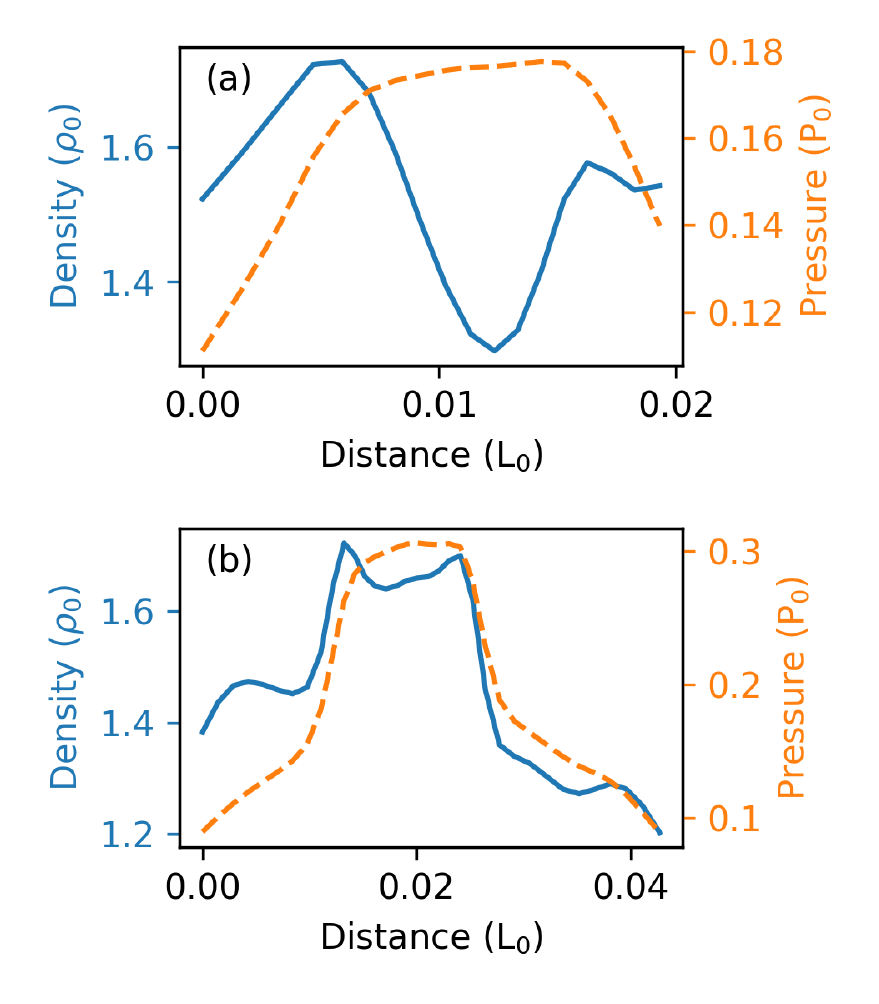}
\caption{Density and gas pressure distributions along the shock front (see red lines on Fig\ref{fig:conduction}(c)(d)) on the post-shocked sides. Panel (a) is relative to Figure \ref{fig:conduction}(c) and panel (b) is for Figure \ref{fig:conduction}(d), respectively. The horizontal axis is the distance measured from the left-end of the shock front. \label{fig:post_var}}
\end{figure}

In Figure \ref{fig:post_var}, we plot density and pressure profiles on the downstream side along the shock front as displayed by red curves in Figure \ref{fig:conduction}. As shown by the cyan lines in Figure \ref{fig:post_var}, obvious density gradients exist along the shock front. The maximum density can be as high as $1.3 \sim 1.4$ times of the minimum density at these two particular times. In the case with thermal conduction (Figure \ref{fig:post_var} (b)), the pressure profile shows a higher center bump, which matches well with density distribution because thermal conduction causes a more temperature diffusion along the TS front. Except the significant density enhancement at the center, similar to the case without thermal conduction, the left side of the shock is, in general, slightly denser than the right side, which is largely due to the asymmetry of the termination shock. Interestingly, in the radio spectral imaging observations made by \citet{Chen2015}, there does seem to a density gradient along the TS front in their Fig. 3(A):  the radio frequency distribution of the radio sources, which is a measure of the local plasma density, is not constant along the shock front, but instead shows a systematic decrease from left to right. Although the observed density gradient may be attributed to a different origin such as the projection effect, and a detailed comparison is beyond the scope of our current paper (largely due to the 2D nature of the simulation and observation), we point out that the dynamic evolution of TSs can undoubtedly introduce a density variation along the TSs, which may be compared with future observational results of TSs made with radio spectral imaging .  
%
%

\begin{figure}[h]
\plotone{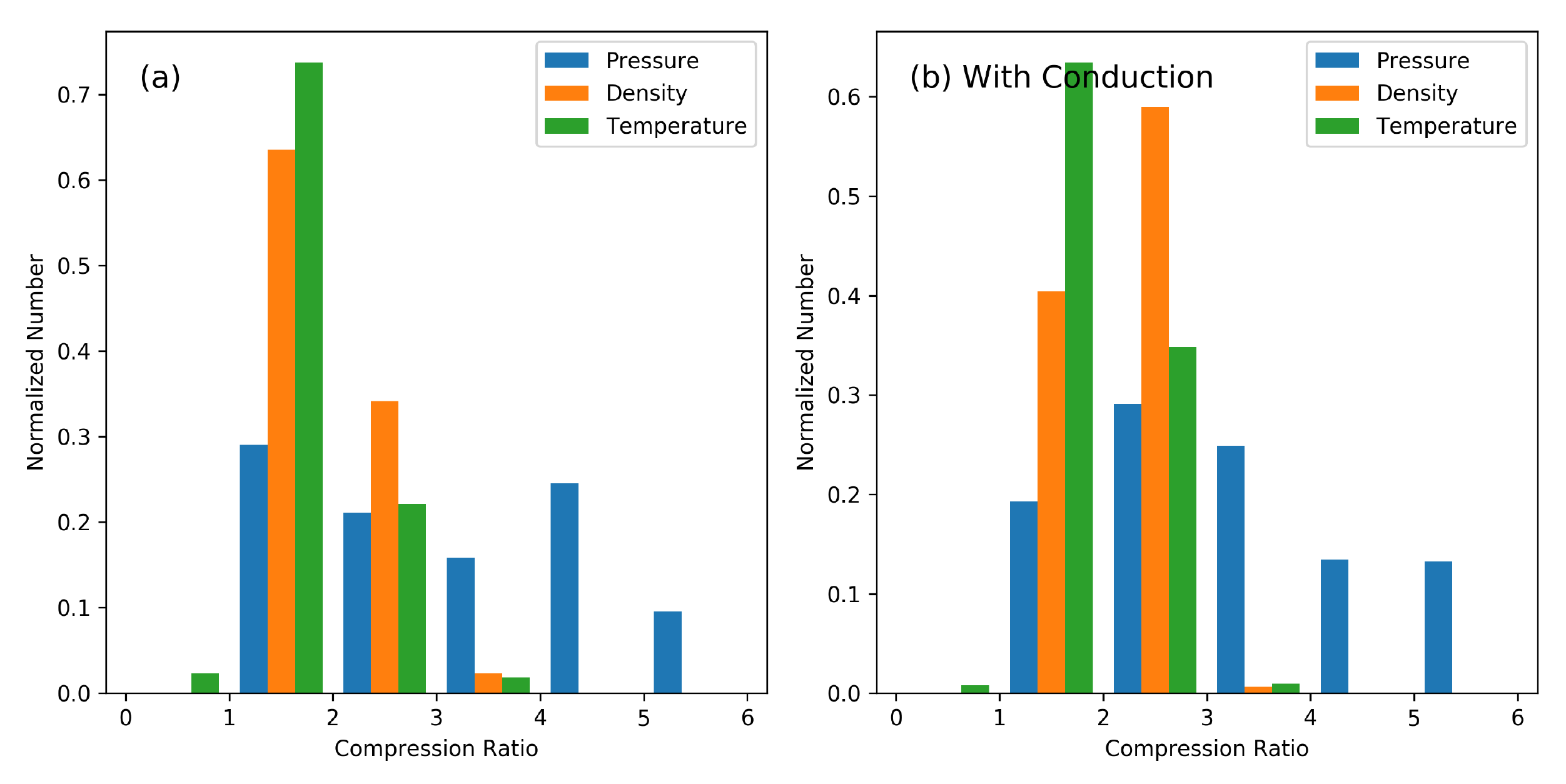}
\caption{Compression ratio distribution in two cases with and without thermal conduction.  \label{fig:ratio_dist_thermal}}
\end{figure}

We perform statistical analysis during a short period and compare the shock compression ratio distribution between two cases in Figure \ref{fig:ratio_dist_thermal}. 
We include a set of sampling points along the TS front at each time within $\sim$2 Alfv\'en times around these two particular times shown in Figure  \ref{fig:conduction}. These sampling points are chosen along the TS front as shown in Figure \ref{fig:divv_2d}. We then interpolate primary variables on both upstream and downstream sides at each sampling point along the local TS normal direction, and obtain pressure, density and temperature ratios across the TS. The normalized number distribution for different compression ratios is plotted in Figure \ref{fig:ratio_dist_thermal}.
We can see that the gas pressure ratio can be higher than 5 in both cases, and density ratio can be higher than 3. The dominant density ratio range is from 1 to 3, and dominant temperature ratio is less 2 in both cases. The basic features are similar between these two cases. However, the higher density compression ratio ($>2$) occurrence probability is about 60\% for the thermal conduction case and less then $\sim$35\% in Case 7. Temperature ratios larger than 2 are slightly more frequent in thermal conduction case.

\section{Implications}
%
%

Simulations performed in this work revealed a dynamical nature of the flare TS driven
by intermittent fast outflows of magnetic reconnection. Here we discuss briefly the implication of the above results on the acceleration of particles at the TS and observations of TS signatures.

As the upstream plasmoids and magnetic fluctuations interact with the magnetic loop in the loop-top region, the flare 
TS is likely to be turbulent and rippled, which may enhance the injection rate for electron acceleration \citep{Guo2012,Guo2015b}.  The typical compression ratio of 1.5--2.5 produced in the simulations is consistent with the value reported by \citet{Chen2015}, $C=n_{\rm down}/n_{\rm up}\approx 1.7$ implied from the observed split-band spectral feature associated with the radio emission from the TS, assuming the high- and low-frequency lane are from the downstream and upstream side of the shock respectively. In diffusive shock acceleration theory, the compression ratio implies a power-law distribution of  $f(E) \propto E^{-\delta}$ with  $\delta\sim$2.5--5.0 \citep{Blandford1987}. This seems to be consistent with those derived from observations of coronal hard X-ray (HXR) sources for a large number of flare events at or near the limb, in which the intense footpoint HXR sources are occulted behind the limb and the weaker coronal HXR sources are revealed \citep{Effenberger2017}. The strong variations in compression ratio in space and time should cause rapid variations in the flux and spectrum of accelerated particles. Future particle acceleration studies need to take into account the more realistic shock features. 
%
%
%
%

Magnetic configurations in post-shocked regions also show highly complex features in our numerical experiments. 
The corresponding plasma compression and curved magnetic field lines can be seen in these regions as well due to the interaction between downflows and the loop top (as shown in Figure \ref{fig:blob_ts_min}).
These features suggest that energetic electrons have more opportunity to be trapped in downstream regions. In fact, \citet{Chen2015}'s analysis has shown that radio spike sources corresponding a TS appear to separate movements after the TS has been disturbed by plasma downflows. These source movements are likely to be caused by changes of electron spatial distribution, or magnetic structures that trap energetic electrons downstream. Therefore, further studies of the electron acceleration mechanism across TSs and dynamical evolution properties downstream are needed to compare with radio and hard X-ray observations in detail.  

The dynamical TS evolution might affect observable signals in spectrum profiles for spectral lines of high temperature plasma. \citet{Guolijia2017} model the synthetic emission of the Fe XXI 1354.08 \AA\ line along the reconnection outflow(and downflow) direction, which is sensitive to $\sim$10 MK plasma and is routinely observed by the  Interface Region Imaging Spectrograph (IRIS). They inserted an artificial termination shock into a Petschek-type reconnection geometry with plasmoid instability, and found that the termination shock leads to enhanced line intensity as well as blue- and/or red-shifted features (depending on the viewing geometry) in the Fe XXI line profile, which are mainly associated with the heated and compressed plasma 
in the downstream of the TS. They suggested such signatures  may be identified in the observed IRIS Fe XXI spectra. As our numerical simulations show, the dynamical TS causes the highly complex TS front and post-shocked plasma structures. In this case, the Fe XXI line profile may include more abundant fine features in either the line width or emission strength, which is worthwhile for further investigations in the future.
%
%
%
%

\section{Summary}
In this work, we performed a set of numerical experiments within the framework of the classic Kopp-Pneuman flare configuration to investigate the formation and dynamics of TSs in solar flares. 
We focus on the region of the lower portion of the reconnection current sheet and the top of the closed flare arcades where the TS is formed. We find that as the reconnection rate increases, the speed of the reconnection outflow steadily grows and, at some point, exceeds the local magnetosonic speed at the top of the flare arcades, thereby forming a TS at the looptop. We further show that the TS is highly dynamic, which responses promptly to the interaction between the intermittent reconnection flows in the shock upstream region and the obstacle in the downstream---the closed flare arcades. The TS front can be clearly delineated by the minimum of the velocity divergence, which allows us to analyze the physical parameters along the TS front in detail. To the contrary of previous assumptions for a steady-state, flat TS, however, our simulations show that the TS front can be highly variable and asymmetric, displaying a variety of morphologies from flat, sloped, curved, to fragmentary or completely disrupted during the arrival of plasmoids of different sizes. We also make detailed measurements along the TS front on its compression ratio, fast mode Mach number, and inclination as it evolves. We find that, although they vary significantly with time and location, the fast mode Mach number and the density compression ratio ranges in $M_F\approx$1--3 and $C\approx$1.5--2.5, respectively, which are broadly consistent with the results derived from the radio observations by \citet{Chen2015}. We also investigate the effects of reconnection guide field and anisotropic thermal conduction on the shock formation and strength. We find that a strong guide field would, in general, reduce the shock strength or even suppress the shock formation completely. The introduction of thermal conduction would alter the details of the density compression and temperature distribution along the shock, and may widen the reconnection outflows and in turn, the width of the TS, but it has relatively small impacts on the shock Mach number. 

Our main findings are summarized as follows:
%
%
%
%

%
%

1. The morphology of the TS varies with time. During a short quasi-steady period, the TS front appears as the combination of one ideal flat shock front and two oblique components. Once the shock has been disturbed by enhanced outflows, downward moving plasmoids or oscillation of loop top structures, the shock shape becomes highly dynamical and the variation can quickly take place in a few of Alfv\'en times scale. The typical length of TS front profile temporally varies in the range of $<0.01L_0$ to $0.09L_0$ ($<1$ to $6.8$ Mm), which is consistent with the previous observations (e.g., $\sim$5 Mm in \cite{Chen2015}).  
%
%

2. The TS front normally consists of both perpendicular and oblique fast-mode shocks. The location of the perpendicular shock is not always at the system center. The slope of the shock front in the perpendicular region can change away from horizontal. However, the center region of a TS front appears to be nearly perpendicular, where the interior angle between shock normal and magnetic field direction is close to 90 degrees. Quasi-parallel shock fronts are rare, and only can be seen at the edge of most compression regions where the shock compression is relatively weak.  

3. A high density compression ratio can be found in dynamical TSs, and ranges from 1.5 to 2.5 for the most strongest compression regions. The maximum compression ratio can be as high as 3.7 at some particular regions and times.
The density compression ratio implied in \citet{Chen2015} is $\sim$ 1.7 obtained from the bandsplitting feature, which may correspond to a weak-shock case in our simulations. Along the TS front, the density variation is also easy to see due to the dynamical evolution, which could be a observational feature in future works. 
Pressure and temperature ratios also show highly dynamical features. The temperature ratios are 2 $\sim$ 3, and the pressure ratio has normally larger values ranging from 1 to 5, and even close to 10.
%
%
%
%

4. Downward moving plasmoids can significantly reduce the strength of the termination shock. A well developed plasmoid may totally destroy the previous TS front, but a new one appears once the plasmoid merges into closed flare magnetic loops. This process leads to violent oscillation of TS height. That is an important feature of intermittent magnetic reconnection. 

5. A background magnetic guide field causes both lower Mach numbers and compression ratios for TSs. A strong guide field obviously increases the local Alfv\'en speed, which results in lower Mach number. However, the impact of a weak guide field is relatively small. In cases with 5\% or 25\% guide field, both the Mach number and compression ratio can still be larger than 2. 

The basic features of TSs in simulations with anisotropic thermal conduction are consistent with other cases that do not include it. Because the unsteady magnetic reconnection rate is not strongly affected by thermal conduction, it is not surprising that the maximum Mach number is around $M \sim 2$, which is in the same range as cases without thermal conduction. The compression ratio distributions are also similar with other cases. However, the thermal conduction can lead to wider reconnection outflows. Therefore, the length of TSs can be slightly larger.

Finally, the findings of the paper need to be further confirmed and extended in three-dimensional (3D) simulations. Recent large-scale 3D numerical simulations have shown that the reconnection outflow region is filled with 3D magnetic structures and turbulence \citep{Daughton2011,Guo2015,Huang2016,Beresnyak2017}. The shock front is likely to be fully turbulent and rippled at a range of spatial scales. We defer a detailed analysis and discussion on 3D simulations to a future study.

\acknowledgments
The authors thank Nicholas A. Murphy and Jing Ye for valuable suggestions that helped to perform MHD simulations and improve this paper. 
This research was supported by National Science Foundation grants AGS-1723313, AGS-1723425 and AST-1735525 to the Smithsonian Astrophysical Observatory.
X.K. acknowledges the support from the National Natural Science Foundation of China under grants 11873036, 11503014 and 11790304 (11790300), and the Young Scholars Program of Shandong University, Weihai. 
F.G. acknowledges the support from the National Science
Foundation under grant No. 1735414 and support from by the U.S.
Department of Energy, Office of Science, Office of Fusion
Energy Science, under Award Number DE-SC0018240. 
B.C. acknowledges the support by NASA grant NNX17AB82G, NSF grants AGS-1654382 and AST-1735405 to the New Jersey Institute of Technology.
The simulations are performed on Hydra cluster operated by Smithsonian Institution using Athena code \footnote{https://princetonuniversity.github.io/Athena-Cversion/}.




\end{document}